# Nanofabricated and integrated colour centres in silicon carbide with high-coherence spin-optical properties

Authors:

Charles Babin[1], Rainer Stöhr[1], Naoya Morioka[1,2], Tobias Linkewitz[1], Timo Steidl[1], Raphael Wörnle[1], Di Liu[1], Erik Hesselmeier[1], Vadim Vorobyov[1], Andrej Denisenko[1], Mario Hentschel[3], Christian Gobert[4], Patrick Berwian[4], Georgy V. Astakhov[5], Wolfgang Knolle[6], Sridhar Majety[7], Pranta Saha[7], Marina Radulaski[7], Nguyen Tien Son[8], Jawad Ul-Hassan[8], Florian Kaiser[1,*], Jörg Wrachtrup[1]

Affiliations:

[1] 3rd Institute of Physics, IQST, and Research Centre SCoPE, University of Stuttgart, 70569 Stuttgart, Germany

[2] Institute for Chemical Research, Kyoto University, Uji, Kyoto, 611-0011, Japan

[3] 4th Institute of Physics, IQST, and Research Centre SCoPE, University of Stuttgart, 70569 Stuttgart, Germany

[4] Fraunhofer Institute for Integrated Systems and Device Technology IISB, Schottkystr. 10, 91058 Erlangen, Germany

[5] Helmholtz-Zentrum Dresden-Rossendorf, Institute of Ion Beam Physics and Materials Research, 01328 Dresden, Germany

[6] Department of Sensoric Surfaces and Functional Interfaces, Leibniz-Institute of Surface Engineering (IOM), 04318 Leipzig, Germany

[7] Department of Electrical and Computer Engineering, University of California, Davis, CA 95616, USA

[8] Department of Physics, Chemistry and Biology, Linköping University, SE-58183 Linköping, Sweden

Corresponding author:

* f.kaiser@pi3.uni-stuttgart.de

Abstract:

Optically addressable spin defects in silicon carbide (SiC) are an emerging platform for quantum information processing. Lending themselves to modern semiconductor nanofabrication, they promise scalable high-efficiency spin-photon interfaces. We demonstrate here nanoscale fabrication of silicon vacancy centres ($V_{Si}$) in 4H-SiC without deterioration of their intrinsic spin-optical properties. In particular, we show nearly transform limited photon emission and record spin coherence times for single defects generated via ion implantation and in triangular cross section waveguides. For the latter, we show further controlled operations on nearby nuclear spin qubits, which is crucial for fault-tolerant quantum information distribution based on cavity quantum electrodynamics.

## Introduction:

Quantum information distribution enables provably secure communication[1], blind quantum computing[2] and advanced quantum metrology[3]. Large-scale quantum networks can be realised based on a well-orchestrated interplay between stationary and flying multi-qubit systems[4]. In this regard, by addressing atom-like transitions in optically active solid-state spins[5,6], impressive landmark demonstrations showed multi-node quantum networks[7], quantum error correction[8] and entanglement distillation[9].

To claim system scalability, the field has yet to improve the interaction between colour centres and photons. E.g., the emission of an optical dipole usually covers a solid angle of $4\pi$, and the associated low light collection efficiency results in slow experimental rates[10], reduced single-shot readout fidelity, and limited quantum protocol complexity. Near-deterministic light-matter interaction is achieved via colour centre integration into monolithic resonators[11–13]. A challenge remains to preserve spin-optical coherence while granting high-fidelity access to qubit clusters, e.g., nuclear spins. The nitrogen vacancy centres in diamond[14] and divacancy spins in SiC[15] showed excellent control over nuclear spins. However integration into nanophotonics structures proved challenging as spin-optical coherences are degraded by coupling to nearby spins and charge traps, which can be present in the starting material and/or are formed during nanofabrication[13,16–21]. Group-IV defects in diamond show robust symmetry-protected optical coherences[22,23], but high-fidelity spin manipulation requires millikelvin temperatures at which direct control over nuclear spin clusters becomes challenging[24,25].

By contrast, deep-bulk $V_{Si}$ centres in 4H-SiC have shown excellent optical coherences up to $T = 20$ K due to wavefunction symmetry protection[26,27], with millisecond spin-coherence times[28] and coherent coupling to nuclear spins[29]. This led to the demonstration of a coherent light-matter interface comprising one electron spin and two photons, suitable for cluster state generation[30]. The single dipole orientation of $V_{Si}$ centres facilitates integration into nanophotonic resonators, however, little is known about spin-optical coherences in such structures.

Here, we demonstrate that $V_{Si}$ centres retain excellent spin-optical properties after generation via helium (He) ion implantation, which allows for three-dimensional positioning. Compared to previous approaches[31–34], we use low-energy ions to create shallow $V_{Si}$ centres with a yield of 8%. We show nearly lifetime limited absorption lines and record Hahn echo spin-coherence times. Further, we simulate SiC waveguide designs[35] and create triangular cross-section nanophotonic waveguides via reactive ion etching (RIE). For $V_{Si}$ centres in such waveguides, we report nearly lifetime limited absorption lines without degradation of spin coherences compared to defects in bulk material. Capitalising on those advances, we coherently control nearby nuclear spins with near unity fidelity, which represents a significant step forward towards integrated multi-spin-multi-photon clusters.

## Results:

We investigate the cubic lattice site silicon vacancy centre (V2) in 4H-SiC, see Fig. 1a. V2 centres are characterised by a zero-phonon line (ZPL) at 917 nm, with a Debye-Waller factor (DWF) of 8-9%.[26,36] As shown in Fig. 1b, the system is spin-$\frac{3}{2}$, in which the sublevels $m_S = \pm\frac{1}{2}$ and $m_S = \pm\frac{3}{2}$ are degenerate Kramers doublets[37]. Optical transitions are spin-conserving, resulting in two absorption lines, labelled A₁ and A₂, respectively[38].

As shown in Fig. 1c, we generate arrays of V$_{Si}$ centres via implantation of He⁺ ions. They are sent through a polymethyl methacrylate (PMMA) mask with 100 nm diameter holes, lithographed on the a-face of an epitaxially grown SiC sample with a low nitrogen concentration in the $[\text{N}] = 4 \cdot 10^{13}$ cm$^{-3}$ range (for more details, see Methods). Compared to previous work[31–34,39,40], we choose a low He⁺ ion energy of 6 keV to minimise crystal damage. To remove residual lattice damage, we subsequently anneal the sample in argon atmosphere at 600°C for 30 minutes. Fig. 1d shows the room-temperature photoluminescence map of an implanted array using off-resonant excitation at 785 nm. We infer the defect number per spot using autocorrelation measurements and obtain a Poissonian distribution with a mean value of $0.67 \pm 0.06$ V2 centres per spot. Considering the ion dose of $1 \cdot 10^{11}$ cm$^{-2}$, we determine an implantation yield of $8.5 \pm 0.8\%$, similar to previous work[34]. From the fluorescence map, we infer that the implanted defect centres are laterally distributed with a variance of $\pm 54$ nm (one standard deviation), which is mainly determined by the implantation mask hole size. Synopsys Sentaurus Monte Carlo simulations show that V2 centres are created about 30-40 nm beneath the surface without channelling effects (see Supplementary Material).

Although high-accuracy shallow defect generation is crucial for integration into nanophotonic resonator devices and for quantum sensing applications, it remained unclear whether spin-optical coherences can be maintained after ion implantation[32,34]. For many colour centres, ion implantation and surface proximity leads to strong degradation of spin-optical properties[41–43]. Here, we show that V2 centres do not suffer from those aspects.

To benchmark the quality of the resonant absorption linewidths, we first measure the excited state lifetime of a single V2 centre at a temperature of $T = 10$ K with a 780 nm ps-pulsed pump laser. The data in Fig. 1e reveals a single-exponential decay with a 1/e-lifetime of $\tau = 7.08 \pm 0.06$ ns, similar to previous work[39]. Thus, the lifetime limited absorption linewidth is $\Delta\nu = (2\,\pi\,\tau)^{-1} = 22.5 \pm 0.2$ MHz. Additional data for proton and silicon ion implantation are given in the Supplementary Material.

Subsequently, we perform resonant excitation with a tunable diode laser at 917 nm (Toptica DL pro). We provide an additional microwave (MW) drive at 70 MHz to continuously mix the ground states[29], this way counteracting optical spin-pumping[38]. A typical resonant photoluminescence excitation scan is shown in Fig. 1f. The two absorption lines show the well-known separation of 1 GHz.[26] The fitted linewidths are $\Delta\nu_{A_1} = 41 \pm 2$ MHz and $\Delta\nu_{A_2} = 24 \pm 1$ MHz for the A₁ and A₂ transitions, respectively, which is close to the lifetime limit. A long-term absorption line measurement is shown in Fig. 1g, revealing little drift with no signs of ionisation. Averaging over 125 scans, we find a linewidth of 40 MHz (29 MHz) for the A₁ (A₂) transition, with a distribution (one standard deviation) of 10 MHz (5 MHz), respectively. Those measurements corroborate the excellent stability of the V2 centre optical lines. We highlight that these results are reproducible for other V2 defects (see Supplementary Material).

We now proceed to V2 centres in nanofabricated waveguides[44,45]. We focus on triangular cross section waveguides, with which high cooperativity cavities have been realised[46]. Additionally, this geometry is very versatile and can be directly applied to bulk substrates of any polytype, without the need for doping level engineering[13]. In colour centre integration with triangular waveguides, single-mode light propagation is crucial for quantum applications. The relationship between device profile and its supported mode wavelength has been modelled recently[35]. Here, we expand on those findings and simulate light propagation in triangular cross section waveguides with a half-opening angle of $\alpha = 45°$ (see Supplementary Material). Fig. 2a-b show the fundamental transverse electric (TE) modes that are supported by waveguides with widths of $d = 400$ nm and $d = 1000$ nm, respectively. Fig. 2c shows the coupling efficiency of light emitted by a centrally located horizontally polarised dipole at 917 nm into the fundamental transverse electric (TE) mode as a function of $d$. For $d = 500$ nm, we find high light guiding efficiencies up to 82% (41% in each propagation direction). The $d = 1000$ nm device still shows a good coupling efficiency of 30% while some of the out-scattered fluorescence can be detected from the top of the waveguide with our collection optics.

To ensure that the V2 centre dipole is parallel to the top surface of the waveguides and to highlight reproducibility across different SiC crystals, we now use a different epitaxially grown SiC layer. The sample was grown along the a-side with a slightly higher nitrogen concentration, $[N] \sim 3 \cdot 10^{15}$ cm$^{-3}$. Prior to nanofabrication, we create V2 centres throughout the entire sample at a density of 0.4 μm$^{-3}$ using electron irradiation at 2 MeV with a dose of 2 kGy (fluence of approximately $5 \cdot 10^{11}$ cm$^{-2}$).

To fabricate waveguide devices, we decide against photoelectrochemical etching, which may deteriorate spin-optical properties due to material degradation and porosification[13]. As shown in Fig. 2d, we fabricate triangular cross-section waveguides via a two-stage process based on RIE with SF$_6$ gas. An initial straight etch creates ~2 μm deep trenches. Subsequent angled etching in a 45° Faraday cage results in suspended waveguide structures with a length of 20 μm and widths of 400 nm and 1000 nm, respectively. A scanning electron microscope image of the latter ones is shown in Fig. 2e.

A typical photoluminescence map with optical excitation and collection from the top of the waveguides is shown in Fig. 2f, revealing multiple bright spots along the waveguides. Emission spectra reveal most of the spots as surface-related or unknown defects (see Supplementary Material). Bright single V2 centres are found in one out of five waveguides with 1000 nm width. Intuitively, this compares low to the average V2 density in the bulk. However, our simulations show that only emitters at the waveguides' top apices show a sizable emission towards our collection optics. The fluorescence of most emitters is actually guided along the waveguides or out-scattered towards the bottom (see Supplementary Material). In analogy to the measurements on implanted V2 centres, we perform resonant photoluminescence excitation scans. Fig. 2g shows a one-hour long recording, revealing narrow absorption lines with no signs of ionisation. Considering that we primarily investigate defects at the waveguide apices, we assign the remaining slow drift to surface charge fluctuations.

To demonstrate that the drift is slow enough to perform complex long-term measurements, we stabilise the resonant excitation laser on optical transitions via software-control[47] (see Supplementary Material). Fig. 2h shows absorption line scans repeated over one hour with the laser feedback activated once per minute and re-centring the scanning window after every scan with respect to the position of the A$_1$ transition. Little to no effective drift is observed[19,41], which enables long-term measurements. To highlight reproducibility, we show the resonant absorption spectra of V2 centres in six different waveguides in Fig. 2i. We find that three out of six V2 centres show nearly lifetime limited absorption lines. The other three still present spin-selective optical transitions. The best defect shows a linewidth of 34 MHz (22 MHz) for the $A_1$ ($A_2$) transition, after averaging over 198 scans. The standard deviation in these measurements is 8 MHz (5 MHz). Unfortunately, we did not identify V2 centres in the 400 nm wide devices. We attribute this to the single-mode operation of the 400 nm waveguides, which suppresses out-scattering towards our collection optics, thus requiring dedicated output couplers[22,48].

The next step should combine nanofabrication and localised defect generation, for which multiple strategies exist. E.g., waveguide fabrication around pre-characterized implanted defect arrays, or depositing He$^+$ ion implantation masks on produced waveguides. For the 45° etched waveguides, both strategies require precise alignment in the range of a few 10 nm. Our simulations indicate a reduced demands on alignment in single-mode devices with a wider opening angle (75° and 900-1000 nm width, see Supplementary Material). Such devices could be created with adapted Faraday cages. Based on the success in the diamond platform[11,46], we believe that mask-free implantation would be ideal, which can be achieved using a helium Focussed Ion Beam source or a helium ion microscope operated at fluxes of about 10 He$^+$ ions per spot.

Having demonstrated implantation and nanofabrication with minimal degradation of the optical properties of V2 centres, we proceed to spin coherence measurements. To maximise spin contrast, we use resonant excitation[29] and take advantage of our software-controlled laser feedback system. Intuitively, we expect that spin properties are less affected than optical properties, however, ion implantation and waveguide nanofabrication can still lead to an increased abundance of parasitic spin defects[13,49]. To allow selective addressing of the three ground state spin transitions via MW drive, we lift the Kramers degeneracy by applying an external magnetic field of $B_0 = 36$ G along the crystal's c-axis (see Fig. 3a). We then deterministically initialise the spin state[29], here into $m_S = +\frac{1}{2}$, followed by probing dephasing and coherences between the levels $m_S = +\frac{1}{2}$ and $m_S = +\frac{3}{2}$.

Ramsey interferometry shows a dephasing time of $T_{2,\text{implanted}}^* = 34 \pm 4 \ \mu s$ for the ion-implanted defects (Fig. 3b), comparable to deep-bulk defects in similar SiC crystals[15,29]. Fig. 3c shows the data for V2 centres in waveguides. We find $T_{2,\text{waveguide}}^* = 9.4 \pm 0.7 \ \mu s$, which is only twice shorter compared to deep bulk defects in the same sample, $T_{2,\text{bulk}}^* = 21 \pm 1 \ \mu s$ (see Supplementary Material).

Hahn echo measurements with ion-implanted V2 centres are presented in Fig. 3d, showing a coherence time of $T_{2,\text{implanted}} = 1.39 \pm 0.10$ ms. This is, to our knowledge, the longest-ever reported Hahn-echo coherence for this system. Fig. 3e shows the data for defects in waveguides, resulting in $T_{2,\text{waveguide}} = 0.84 \pm 0.01$ ms, which is comparable to bulk defects in the same sample, $T_{2,\text{bulk}} = 0.85 \pm 0.03$ ms (see Supplementary Material). Thus, RIE can be used for nanofabrication of high-quality SiC material without deteriorating spin properties.

Interestingly, the data in Fig. 3e displays a strong modulation due to dipolar interaction with weakly coupled nuclear spins. This indicates that the hyperfine coupling strength with at least one nucleus is comparable to the Larmor frequency. We show now that the system is described by one V2 centre coupled to two nuclear spins, as depicted in the artistic design in Fig. 4a. To infer the coupling strengths with high resolution, we apply a slightly increased magnetic field, $B_0 = 81$ G and perform 8-pulse Carr-Purcell-Meiboom-Gill (CPMG) sequences with varying waiting time $\tau_{\text{CPMG}}$. From the modulation signal in Fig. 4b, we extract that both nuclei are $^{29}$Si spins with hyperfine coefficients of $A_{\parallel,1} = 2\pi \cdot (-23.5)$ kHz and $A_{\perp,1} = 2\pi \cdot 12.0$ kHz for the first nuclear spin (N$_1$), as well as $A_{\parallel,2} = 2\pi \cdot 0.2$ kHz and $A_{\perp,2} = 2\pi \cdot 8.5$ kHz for the second one (N$_2$, see Supplementary Material).

Having determined the coupling coefficients, we now show coherent manipulation of an electron-nuclear spin pair in a SiC waveguide. Compared to previous work[15,50,51], we implement spin control sequences at a low magnetic field ($B_0 = 81$ G). This keeps the hyperfine coupling strength comparable to the nuclear spin Larmor frequency $\omega_L$, which allows us to perform nuclear spin quantum gates in short times using a moderate number of CPMG pulses.

As the V2 centre ground state is a quartet, it directly provides three exploitable spin subspaces ($m_S = [+\frac{3}{2}; +\frac{1}{2}]$, $m_S = [+\frac{1}{2}; -\frac{1}{2}]$, and $m_S = [-\frac{1}{2}; -\frac{3}{2}]$) that allow to implement vastly different nuclear spin control sequences. Here, we focus on the subspace $m_S = [+\frac{3}{2}; +\frac{1}{2}]$ to demonstrate nuclear spin rotations around the $x$-axis for the first nuclear spin. The strong resonance signal in Fig. 4b at $\tau_{\text{CPMG}} = 5.38 \ \mu s$ already indicates that electron and nuclear spins are flipped. Ensuring precise rotations around the desired $x$-axis requires determining the optimal waiting times $\tau^{(k)}$ (dubbed as resonances), at which the electron spin signal due to the nuclear spin precession is exactly commensurate with the CPMG pulse spacing. To this end, we develop a generalised theoretical model to describe CPMG resonances for weakly coupled nuclear spins[50] in small magnetic fields (see Supplementary Material). For each individual nuclear spin, the resonances occur at

$$\tau_{\text{approx}}^{(k)} = \tau_k \left( 1 + \sin\left( \frac{\omega_{1/2}\,\tau_k}{2} \right) \frac{(\boldsymbol{n}_{1/2} \cdot \boldsymbol{n}_{3/2})^{-1} - 1}{(2k-1)\,\pi} \right).$$

Here, the integer number $k$ indicates the resonance number, $\tau_k = \frac{(2k-1)\,\pi}{\omega_{1/2}+\omega_{3/2}}$, $\omega_{m_S} = \sqrt{(m_S\,A_\perp)^2 + (m_S\,A_\parallel - \omega_L)^2}$ and $\omega_{m_S}\,\boldsymbol{n}_{m_S} = m_S\,A_\perp\,\boldsymbol{e_x} + (m_S\,A_\parallel - \omega_L)\boldsymbol{e_z}$. The unit vectors $\boldsymbol{e_x}$ and $\boldsymbol{e_z}$ denote the nuclear spin rotation axis. For the first resonance, we calculate $\tau_{\text{approx}}^{(1)} = 5.38\ \mu s$, in excellent agreement with the experimental data. We now control the nuclear spin rotation angle by varying the number $N$ of CPMG pulses at $\tau_{\text{CPMG}} = 5.38\ \mu s$. The uncorrected raw data in Fig. 4c reveals that the electron spin signal is modulated due to electron spin-controlled nuclear spin rotations with near-unity fringe contrast, in excellent accord with the simulated signal with no free parameters (see Supplementary Material). By varying the number of refocusing pulses, we implement relevant quantum gates on the bi-partite system. The Bloch spheres in Fig. 4d visualise the electron spin-dependent rotation of the first nucleus $N_1$ for $N = 4$ CPMG pulses. The net result is an opposite rotation around the $x$-axis, which can be used to implement an entangling Hadamard gate. Fig. 4e shows the rotation of the second nucleus $N_2$. The nuclear spin is decoupled as its rotation is (essentially) independent of the electron spin state. By increasing the number of refocusing pulses, we can rotate $N_1$ to implement additional gates, such as the $\sigma_x$-gate for $N = 8$, and the identity operation for $N = 16$. For the three gates (Hadamard, $\sigma_x$, and identity), our simulations show fidelities of 97%, 94% and 98%, respectively (see Supplementary Material). These high fidelities are in part due to the fast implementation of gates using only a small number of refocusing pulses for which electron spin decoherence is negligible. In this sense, we emphasise that the small signal decrease in Fig. 4c for high pulse numbers ($N > 20$) is not associated with electron spin decoherence. The signal decrease stems from a minimal remaining spin-dependent rotation of $N_2$, which can be suppressed with stronger external magnetic fields[7,15]. Our results underline that waveguide-integrated V2 centres can be used for controlling nuclear spins with high fidelities at cryogenic temperatures, promising implementation of quantum memories, quantum error correction, and (distributed) quantum computational tasks.

## Conclusion:

We demonstrated that V2 centres in 4H-SiC can be successfully integrated into nanofabricated waveguide devices produced by RIE techniques. We showed further that low-energy He[+] implantation creates shallow V2 centres with high yield and good spatial accuracy. Despite using SiC crystals from completely different growth processes, we observed nearly lifetime-limited absorption lines and reported record Hahn echo electron spin coherence times. The fact that ideally oriented V2 centres in novel a-plane 4H-SiC can be used for quantum applications offers a unique potential to further scale up semiconductor quantum technologies[44].

To induce these steps, we outlined how to combine ideal photonic waveguide designs and implantation strategies that could be used toward realizing cavity-based spin-photon interfaces with high cooperativities[11]. For maximum coupling efficiency, such devices could be directly connected to optical fibres[52].

To interfere multiple emitters, Stark shift tuning[53] could be used for both matching resonance frequencies and shaping optical linewidths[54]. On-chip wavelength conversion to the telecom band could be implemented by utilizing the $\chi^{(2)}$ and $\chi^{(3)}$ nonlinearities of 4H-SiC[55], ultimately enabling large-scale memory-enhanced quantum repeater networks[46].

To scale-up such networks, nuclear spin quantum memories and processors represent a critical technology[4]. In this sense, we demonstrated coherent control over individual nuclear spins, and we implemented relevant quantum gates with high fidelity. The amount of controllable nuclear spins can be directly scaled up in isotopically engineered samples[15]. In this regard, V2 centres are very promising central spins as they can be operated at temperatures up to $T = 20$ K before experiencing spin-optical coherence degradation[26]. The high cooling powers of standard cryogenic equipment at those temperatures make it possible to operate experiments at full duty cycle while controlling nuclear spins with high-power radiofrequency drive.

Overall, these results corroborate that V2 centres in SiC are attractive for developing large-scale quantum networks based on integrated quantum computational clusters with efficient spin-photon interfaces.

## Data availability

Source data are provided with this paper and at https://doi.org/10.18419/darus-2107. Any further data are available from the corresponding author upon request.

## Code availability

Fits to the data were made using python software. The fit functions, parameters, and simulations of hyperfine interactions can be made available upon request.

## Acknowledgements:

C.B., F.K. and J.W. thank Arnold Weible, Rolf Reuter, Annette Zechmeister, Jonas Meinel, Roland Nagy, Jonathan Körber, Marcel Krumrein, Matthew Joliffe, Matthias Niethammer, Izel Gediz, Jonas Zatsch, Durga Dasari, Yu-Chen Chen, Kai-Mei Fu, Daniil Lukin, and Jelena Vučković for experimental assistance and fruitful discussions. We further acknowledge technical support from Swabian Instruments GmbH and Toptica Photonics AG.

G.V.A. thanks Ulrich Kentsch for the proton implantation assistance and support from the Ion Beam Center at Helmholtz-Zentrum Dresden-Rossendorf (HZDR).

N.T.S. acknowledges the Swedish Research Council (Grant No. VR 2016-04068).

J.U.H. acknowledges the Swedish Energy Agency (Grant No. 43611-1) and Swedish Research Council (Grant No. 2020-05444).

N.T.S. and J.U.H. thank the EU H2020 project QuanTELCO (Grant No. 862721) and the Knut and Alice Wallenberg Foundation (Grant No. KAW 2018.0071).

S.M. acknowledges support by the UC Davis Summer GSR Award.

M.R. acknowledges support by the National Science Foundation under the grant CAREER-2047564.

J.W. acknowledges the EU-FET Flagship on Quantum Technologies through the project ASTERIQS (Grant Agreement No. 820394), the European Research Council (ERC) grant SMel, the Max Planck Society, and the German Research Foundation (SPP 1601, FOR 2724). F.K. and J.W. acknowledge support by the EU-FET Flagship on Quantum Technologies through the project QIA (Grant Agreement No. 820445), as well as the German Federal Ministry of Education and Research (BMBF) for the project Q.Link.X (Grant Agreement No. 16KIS0867).

## Author contributions:

C.B., and F.K. conceived the experiments. F.K. and J.W. directed the research. R.S., T.L., R.W. and A.D. produced masks and performed helium ion implantation. C.G. and P.B. performed helium implantation simulations. R.S., T.L., R.W. and G.V.A. produced masks and designed proton implantation. M.H. performed silicon ion implantation. W.K. performed electron irradiation. C.B., R.S. and T.L. produced SiC waveguides. V.V., S.M., P.S. and M.R. simulated beam profiles in waveguides. N.T.S. and J.U.H. grew the SiC samples. C.B., N.M., T.L., T.S., R.W., D.L., E.H. and F.K. performed the experiments. C.B., N.M., T.L., T.S., D.L. and F.K. developed and improved software. C.B., R.W. and F.K. analysed the data. C.B. and F.K. developed the theoretical framework for nuclear spin control. C.B., M.R., F.K. and J.W. wrote the manuscript. All authors provided helpful comments during the writing process.

## Methods:

### Sample preparation:

We use two different samples for the implantation and waveguide experiments.

For the implantation experiments, we use a 110 µm thick 4H-$^{28}$Si$^{12}$C silicon carbide layer that was grown by chemical vapour deposition (CVD) on the c-plane of a n-type (0001) 4H-SiC wafer. The isotope purity is estimated to be [$^{28}$Si] > 99.85% and [$^{12}$C] > 99.98%, which was confirmed by secondary ion mass spectroscopy (SIMS) for one of the wafers in the series. After chemical mechanical polishing (CMP) of the top layer, the substrate was removed by mechanical polishing and the final isotopically enriched free-standing layer had a thickness of ~100 µm. Current-voltage measurements at room temperatures show that the layer is n-type with a free carrier concentration of ~$6 \times 10^{13}$ cm$^{-3}$. This value is close to the concentration of shallow nitrogen donors of ~$4 \times 10^{13}$ cm$^{-3}$, which was determined by photoluminescence at low temperatures. Deep level transient spectroscopy measurements show that the dominant electron trap in the layer is related to the carbon vacancy with a concentration in the mid $10^{12}$ cm$^{-3}$ range. Minority carrier lifetime mapping of the carrier shows a homogeneous carrier lifetime of ~ 0.6 µs. Since the lifetime was measured by an optical method with high injection, the real lifetime is expected to be double, i.e., ~1.2 µs. Such a high minority carrier lifetime indicates that the density of all electron traps should not be more than mid $10^{13}$ cm$^{-3}$. To investigate colour centres from the a-side of the samples, they were cleaved in order to obtain a reasonably smooth surface morphology. Samples were investigated with fluorescence microscopy and no defect centres were found.

For the waveguide experiments, we used a 28 µm thick 4H-$^{28}$Si$^{12}$C silicon carbide layer that was grown by CVD on an a-plane n-type 4H-SiC wafer. Isotope concentration is estimated to be similar to the above-described c-plane samples. However, we found an increased incorporation of nitrogen donors in the ~$3 \times 10^{15}$ cm$^{-3}$ range, which attributes to slightly increased optical linewidths. The surface roughness of this sample was measured to be below 0.4 nm via an atomic force microscope. Defects in this sample were generated using electron irradiation at 2 MeV with a fluence of 2 kGy. After irradiation, the sample was annealed at a temperature of 600 °C for 30 min to remove some interstitial-related defects.

### Spin manipulation:

For all spin manipulation experiments, MW is provided through a 50 µm diameter copper wire that is placed within 50-100 µm distance to the investigated colour centres. MW signals are generated using signal generators (Rohde & Schwarz SMIQ03b and SMIQ02b) and subsequent switches (MiniCircuits ZASWA-2-50DRA+). MW signals are combined using MW splitters, followed by a 20 dB amplifier stage. Typical MW power levels at the wire are 15-20 dBm.

### He implantation process:

To generate defects via He$^+$ ion implantation, we spin-coated a 200 nm thick 200k-PMMA layer. Electron beam lithography (EBL) was done in Raith Eline apparatus with 20 kV acceleration voltage, 10 µm aperture, and an exposure dose of 270 µC/cm$^2$. After exposure, the sample was placed in a micro-beam implanter (ion gun) with a He$^+$ source and a Wien filter. He$^+$ ions were accelerated at 6 keV and implanted into the sample at a dose of $10^{11}$ ions/cm$^2$. After removing the PMMA mask, the sample was annealed at 600 °C for 30 min in argon atmosphere to remove some interstitial-related defects.

**RIE process:**
To create suspended waveguide structures, a double layer PMMA mask (950K and 200K) was patterned using ELB. A 200 nm thick nickel mask was evaporated onto the surface of the sample. The pattern was transferred onto SiC via SF$_6$ plasma etching process (20 sccm, 7.5 mTorr, -20°C, RIE power of 100 W). After completing straight and angled etching processes, mask residuals were removed by immersion into diluted nitric acid (HNO$_3$). Thereafter, the sample was annealed at 600 °C for 30 min in argon atmosphere to remove some interstitial-related defects.

**Photonic modelling:**
To model the light propagation in SiC waveguides, we use Finite-Difference Time-Domain simulations in Lumerical software. Colour centres are simulated as horizontally oriented point dipoles at 917 nm, placed inside triangular waveguides with $n = 2.6$ refractive index and mesh size of 20 nm. Mode coupling is monitored for six lowest energy modes supported in the waveguide and coupling efficiencies are evaluated after 10 μm of propagation.

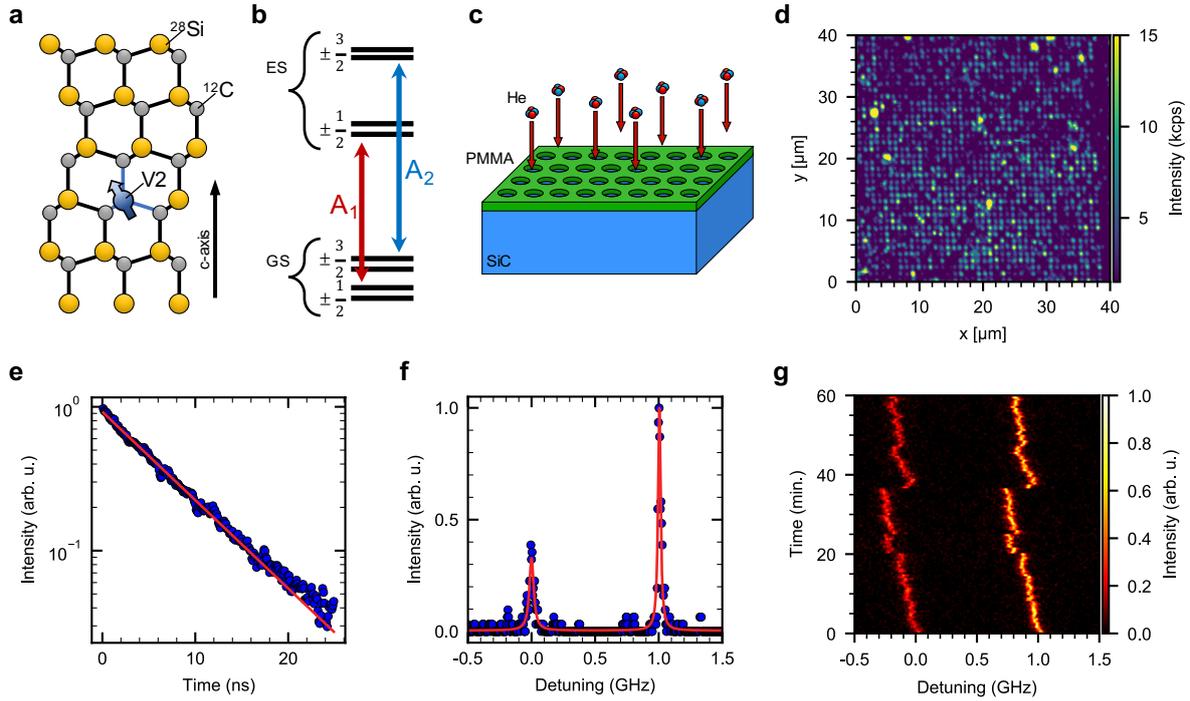

Figure 1: Properties of He$^+$ ion implanted V2 centres. **a** Crystal structure of 4H-SiC. The V2 centre is formed by a missing silicon atom at a cubic lattice site. **b** Energy level diagram of V2 centres. Ground and excited states are spin quartets and optical transitions are spin conserving (A$_1$ − red, A$_2$ − blue). **c** Schematics of V2 centre creation via He$^+$ ion implantation. He$^+$ ions are accelerated to 6 keV and implanted at the a-side of the crystal through a lithographed PMMA mask. **d** Confocal fluorescence scan of the implanted defect centre array. **e** Off-resonant excitation lifetime measurement of a single V2 centre. The fit is based on a single-exponential decay ($\propto \exp\left(-t/\tau\right)$) with a 1/e-lifetime of $\tau = 7.08 \pm 0.06$ ns. **f** Single-scan resonant photoluminiscence excitation scan on an implanted V2 centre. The nearly lifetime limited linewidth confirms the excellent spectral stability of V2 centres. **g** Repeated resonant excitation scans during one hour without repump laser. No ionisation is observed and the small remaining drift is assigned to surface charge fluctuations.

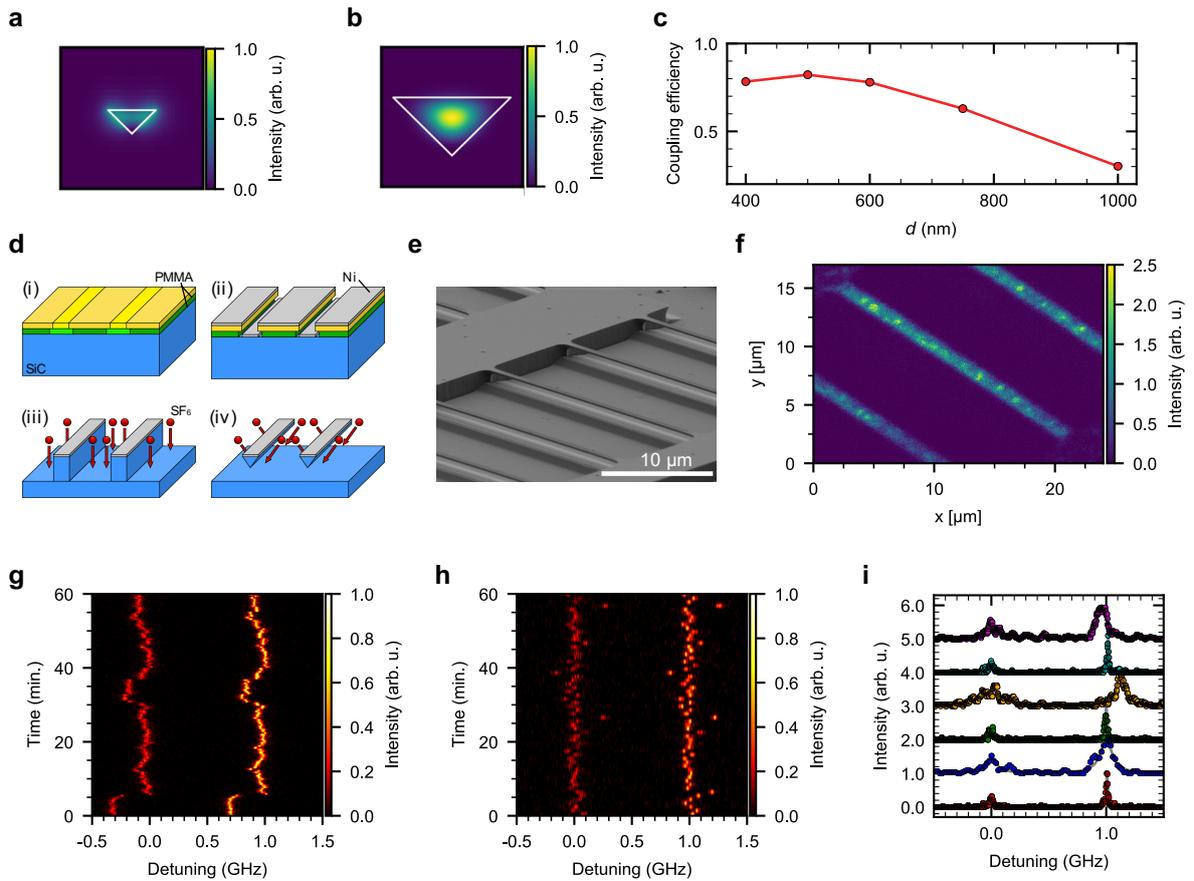

Figure 2: Properties of V2 centres in nanofabricated waveguides. **a** Cross-sectional fundamental TE mode profile in waveguides with a half opening angle $\alpha = 45°$ and width $d = 400\,\mathrm{nm}$. **b** TE mode profile for a waveguide with $\alpha = 45°$ and $d = 1000\,\mathrm{nm}$. **c** Coupling efficiency into the fundamental TE waveguide mode as a function of the width $d$ (emitter assumed to be located in the centre of the waveguide). High efficiencies are reached for $d = 400 - 600\,\mathrm{nm}$. **d** Waveguide fabrication recipe. (i) Deposition of a double layer of PMMA mask; (ii) Deposition of a nickel mask; (iii) A straight reactive ion etch based on $SF_6$ plasma creates $3\,\mathrm{\mu m}$ deep trenches. (iv) Angled $SF_6$ plasma etch creates triangular cross-section waveguides. **e** Scanning electron microscope image of the created waveguide structures that appear smooth and well undercut. **f** Confocal fluorescence microscope image of the waveguides. Bright spots are in majority surface-related defects. In average, one V2 centre is found in every five waveguides with $d = 1000\,\mathrm{nm}$. **g** Resonant excitation scans during one hour. No ionisation is observed and the wavelength drift is minimal. **h** Resonant excitation scans with the laser feedback software activated once per minute. **i** Single-line resonant excitation scans for six different V2 centres in waveguides. For better visibility, the $A_1$ transition for all spectra is centred at zero detuning. The actual spectral distribution of the defects is $\pm 10\,\mathrm{GHz}$, comparable to bulk defects[56].

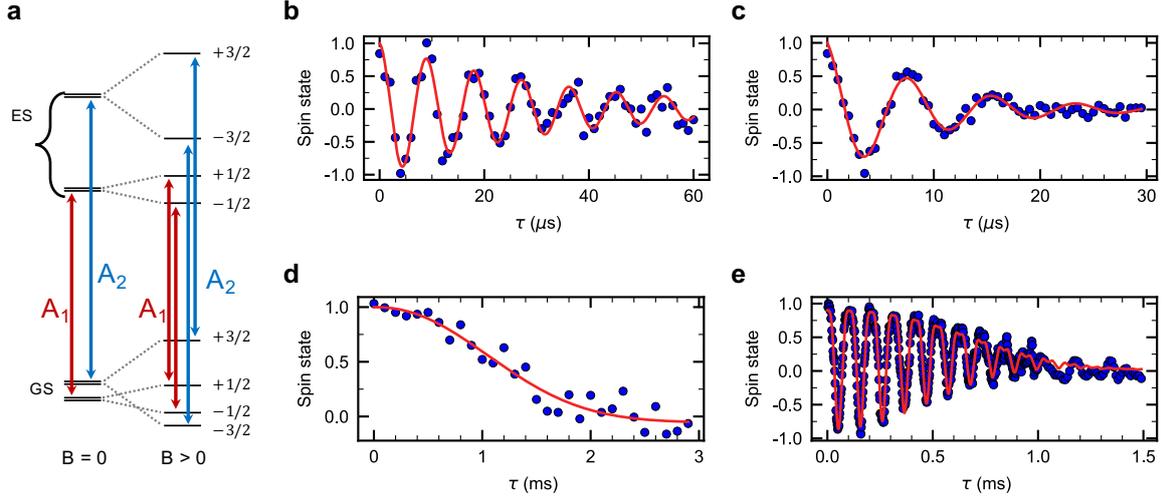

Figure 3: Electron spin properties of nanofabricated V2 centres. **a** Energy level diagram of V2 centres in an external magnetic field to lift the Kramers degeneracy in the ground and excited states. **b** Ramsey interferometry for He⁺ implanted V2 centres. The dephasing time is $T^*_{2,\text{implanted}} = 34 \pm 4 \ \mu s$. **c** Dephasing of V2 centres in waveguides, $T^*_{2,\text{waveguide}} = 9.4 \pm 0.7 \ \mu s$. **d** Spin coherence time of He⁺ implanted V2 centres using Hahn echo. The coherence time is $T_{2,\text{implanted}} = 1.39 \pm 0.06$ ms. **e** Spin coherence in waveguides, $T_{2,\text{waveguide}} = 0.84 \pm 0.01$ ms. The strong modulation corroborates coupling to at least one nearby ²⁹Si nuclear spin. In all plots, spin state +1 corresponds to $m_S = \frac{3}{2}$ and -1 to $m_S = \frac{1}{2}$. The signal envelopes are fitted using stretched exponential functions $\propto A \cdot \exp\left(-\frac{t}{\tau}\right)^N + y_0$.

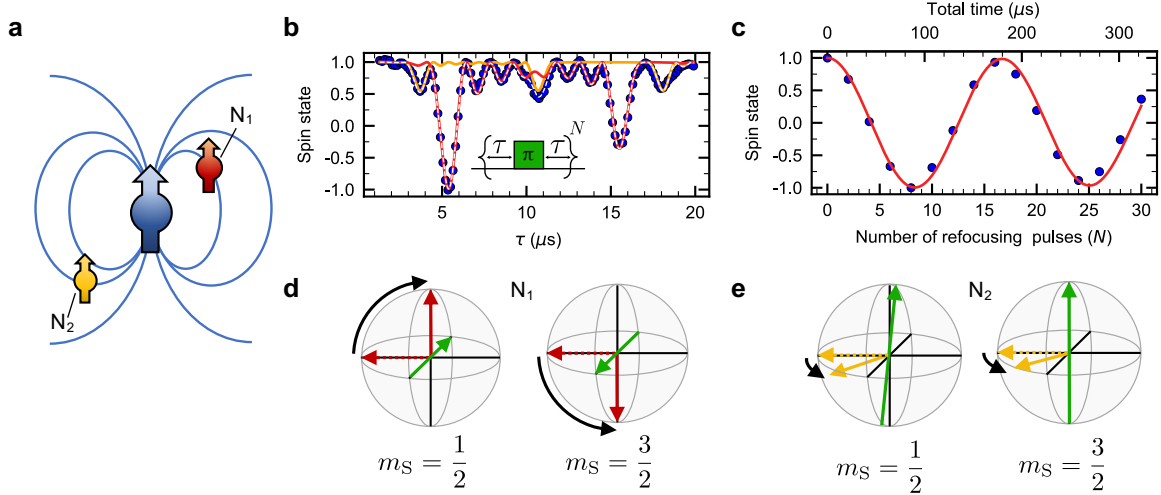

Figure 4: Control over nuclear spin qubits. **a** Artistic representation of the coupled electron-nuclear spin triplet identified in a waveguide. **b** Spin signal for the 8-pulse CPMG sequence (inset) at $B_0 = 81$ G. Blue dots are experimental data. The red and orange lines correspond to the (independent) spin signals obtained for the first and second nucleus, respectively. The white dashed line is the product of both fits. The strongest modulation is observed for N₁ at the resonance time $\tau_{\text{approx}}^{(k)} = 5.38$ μs. **c** Electron-spin signal for different number $N$ of refocusing pulses at $\tau_{\text{CPMG}} = 5.38$ μs. Strong oscillations stem from electron spin-dependent rotations of N₁. Blue dots are uncorrected raw data. The red curve is a simulation based on two nuclear spins without any free parameters. In all plots, spin state +1 corresponds to $m_{\text{S}} = \frac{3}{2}$ and -1 to $m_{\text{S}} = \frac{1}{2}$. **d** Bloch-sphere representation showing the dynamics of the nuclear spin N₁ during the $N = 4$ pulse CPMG sequence. The rotation axes (green arrows) for the electron spin states $m_{\text{S}} = \frac{1}{2}$ and $m_{\text{S}} = \frac{3}{2}$ are antiparallel ($\pm \boldsymbol{e_x}$) with a rotation angle of $\pm \frac{\pi}{2}$ rad. **e** Same as in Fig. 4d, but for the nuclear spin N₂. The nuclear spin is decoupled from the electron spin as the rotation axis is (nearly) the same ($\sim \boldsymbol{e_z}$) with a very small rotation angle of 0.04 rad (the shown rotation is exaggerated for clarity).

# Supplementary information for:
# Nanofabricated and integrated colour centres in silicon carbide with high-coherence spin-optical properties

Charles Babin[1],* Rainer Stöhr[1], Naoya Morioka[1,2], Tobias Linkewitz[1], Timo Steidl[1], Raphael Wörnle[1],
Di Liu[1], Erik Hesselmeier[1], Vadim Vorobyov[1], Andrej Denisenko[1], Mario Hentschel[3], Christian
Gobert[4], Patrick Berwian[4], Georgy V. Astakhov[5], Wolfgang Knolle[6], Sridhar Majety[7], Pranta Saha[7],
Marina Radulaski[7], Nguyen Tien Son[8], Jawad Ul-Hassan[8], Florian Kaiser[1],† and Jörg Wrachtrup[1]

[1] *3rd Institute of Physics, IQST,
and Research Center SCoPE,
University of Stuttgart, 70569 Stuttgart, Germany*
[2] *Institute for Chemical Research,
Kyoto University, Uji, Kyoto, 611-0011, Japan*
[3] *4th Institute of Physics, IQST,
and Research Center SCoPE,
University of Stuttgart, 70569 Stuttgart, Germany*
[4] *Fraunhofer Institute for Integrated Systems and Device Technology IISB,
Schottkystr. 10, 91058 Erlangen, Germany*
[5] *Helmholtz-Zentrum Dresden-Rossendorf,
Institute of Ion Beam Physics and Materials Research, 01328 Dresden, Germany*
[6] *Department of Sensoric Surfaces and Functional Interfaces,
Leibniz-Institute of Surface Engineering (IOM), 04318 Leipzig, Germany*
[7] *Department of Electrical and Computer Engineering,
University of California, Davis, CA 95616, USA*
[8] *Department of Physics, Chemistry and Biology,
Linköping University, SE-58183 Linköping, Sweden*

## CONTENTS

# S1.  V$_{Si}$ IMPLANTATION EXPERIMENTS

In this section, we provide additional information on the different implantation techniques used to generate V$_{Si}$ colour centres. In particular, we compare defect generation via implantation of protons, He$^+$ ions and Si$^{2+}$ ions.

## A.  Defect generation via helium ion implantation

The key results and implantation parameters have already been discussed in the main text. In this section, we provide additional information on experiments that were carried out to infer the implantation yield and accuracy. To underline reproducibility, we also show the resonant excitation spectra of five additional V2 centres.

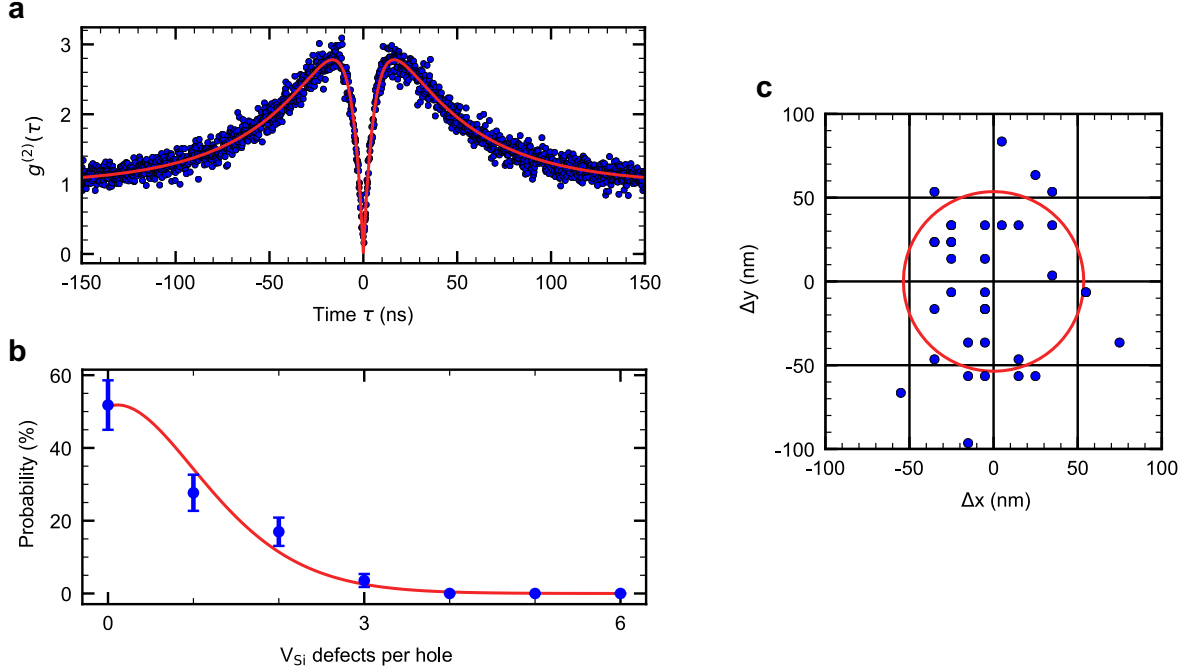

Figure S1.  **a** Typical second order correlation measurement of a single V2 centre. Blue dots are data and the red line is a fit taking into account the below-described function for $g^{(2)}(\tau)$. **b** Probability distribution of the number of V2 centres per implantation site. Blue dots are data and error bars correspond to the square root of the number of measurements. The red line is a fit to the data considering a Poissonian distribution. **c** Lateral spatial distribution of the implanted single V2 centres compared to a simulated grid. The red circle represents the 53.6 nm variance of all positions.

### 1.  Implantation yield

The implantation yield is defined as the number of V2 centres created per implanted ion. To infer the number of defects per implantation spot, we perform a Hanbury Brown and Twiss interferometry. We record the second-order correlation function ($g^{(2)}$) under off-resonant excitation (785 nm with 2 mW pump power). Note that we are only interested in the yield for generating V2 centres. However, we find that V2 and V1 centres [S1] are created at roughly the same rate. To suppress spurious phonon sideband emission of V1 centres during the $g^{(2)}$ measurements, we use a $950 \pm 25$ nm bandpass filter which rejects most of the V1 centre's emission, and, at the same time, shows a high transmission for in the first phonon sideband of the V2 centre emission (see Figure S6**a**).

A typical measurement is shown in Figure S1**a**. In general, the data can be fitted with the function $g^{(2)}(\tau) = \frac{1}{N}[1 - a \cdot e^{-\frac{|\tau|}{\tau_1}} + (1-a) \cdot e^{-\frac{|\tau|}{\tau_2}}] + \frac{N-1}{N}$. Here, $\tau$ is the delay time between two detector clicks in the Hanbury Brown and Twiss setup, $a$ is a prefactor, $\tau_{1,2}$ are decay constants, and $N$ is the number of emitters in a spot. In the measurement shown in Figure S1**a**, we find $g^{(2)}(\tau = 0) \ll 0.5$, which corroborates a single emitter (value extracted from the fit: $N = 1.01 \pm 0.02$). By inquiring 112 spots in the array, we extract the statistical distribution of defect number, shown in Figure S1**b**. The fit to the data is based on a Poissonian function from which we determine the average number of V2 centres created per spot to be $0.66 \pm 0.06$. Considering the ion implantation dose ($1 \cdot 10^{11}$ cm$^{-2}$) and the mask's hole size (100 nm diameter), we obtain therefore an implantation yield of $8.5 \pm 0.8\%$ for generating V2 centres.

### 2. Implantation accuracy

To estimate the spatial implantation accuracy in lateral coordinates, we use a closed-loop piezo scanner (Mad City Labs Inc. Nano-LPMW, 0.4 nm resolution) and determine the position of 36 single V2 centres within a $12 \times 12$ array. Each single-defect position is determined using a two-dimensional Gaussian fit. To infer the principal implantation accuracy, we compare the implantation positions with a simulated grid. We minimize error squares between data and simulation by allowing the simulated grid to account sample rotation and scanner linearity error in both directions. The deviation between the measured data and simulated grid positions is shown in Figure S1**c**. The variance of the defect positions is 53.6 nm (represented by a circle). This value is almost completely determined by the implantation hole size (50 nm diameter), such that the lateral straggle of He$^+$ ions during implantation is minimal. This is confirmed by SRIM simulations for implanting 6 keV He$^+$ ions into SiC that predict a lateral projected straggle of 22 nm. The implantation accuracy at this low energy is thus sufficient for creating defects in the centre of photonic crystal cavities.

### 3. Implantation depth

Simulations of helium ion implantation and corresponding vacancy generation have been carried out using the Synopsys Sentaurus Monte Carlo (MC) simulator in order to take the crystalline structure of 4H-SiC into consideration. Sentaurus MC uses the binary collision approximation (BCA). Implantation was simulated at a temperature of $T = 300$ K into the (1 1 -2 0) plane with an implantation energy of 6 keV and a dose of $1 \cdot 10^{11}$ cm$^{-2}$. As we do not have a precise control on the implantation angle in our implanter device, we perform simulations for a variation of tilt angles between $0°$ and $7°$. For the simulation of vacancy generation, threshold displacement energies of 41 eV for Si and 16 eV for C have been chosen for the Frenkel pair defect model while taking into account all secondary recoils from cascades [S2]. Figure S2**a** shows the results for implantation at an angle of $0°$. Here, we observe a small He$^+$ ion channelling effect for implantation. However, no channelling is found in the depth profiles of carbon vacancies (V$_C$) and especially not for V$_{Si}$. Further simulations at an implantation angle of $7°$, shown in Figure S2**b**, show essentially the shape of the depth profile, however with a $\approx 3\times$ higher defect generation efficiency. Figure S2**c** shows the simulated V$_{Si}$ depth profiles at different implantation angles ranging from $0°$ to $7°$. The general shape of all profiles is the nearly identical, thus confirming that the V$_{Si}$ centres are created in majority at a depth of $30 - 40$ nm.

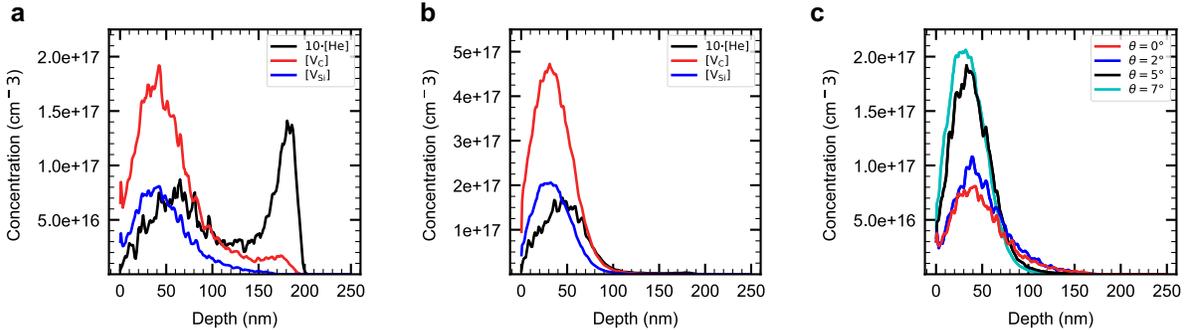

Figure S2. Implantation depths for bombardment with He$^+$ ions at an energy of 6 keV along the (1 1 -2 0) side of 4H-SiC. In both plots, the black line represents the stopping distance for the He$^+$ ions, the green line represents the depths at which carbon vacancies are created (V$_C$) and the blue line shows the V$_{Si}$ depth profiles. **a** Implantation at normal incidence ($0°$). **b** Implantation at an incidence angle of $7°$ with respect to the surface normal. Although the peak of the V$_{Si}$ formation density is $\approx 3\times$ higher in **b**, no significant change in the depth profile is observed. **c** V$_{Si}$ depth profiles at different implantation angles ranging from $0°$ to $7°$.

### 4. Stability of the resonant absorption lines

To confirm that implanted V2 centres reproducibly show a good spectral stability, we show here resonant absorption line measurements on another five implanted defects. For each defect, we show repeated laser excitation scans for 20 minutes. The results are shown in Figure S3**a-e**.

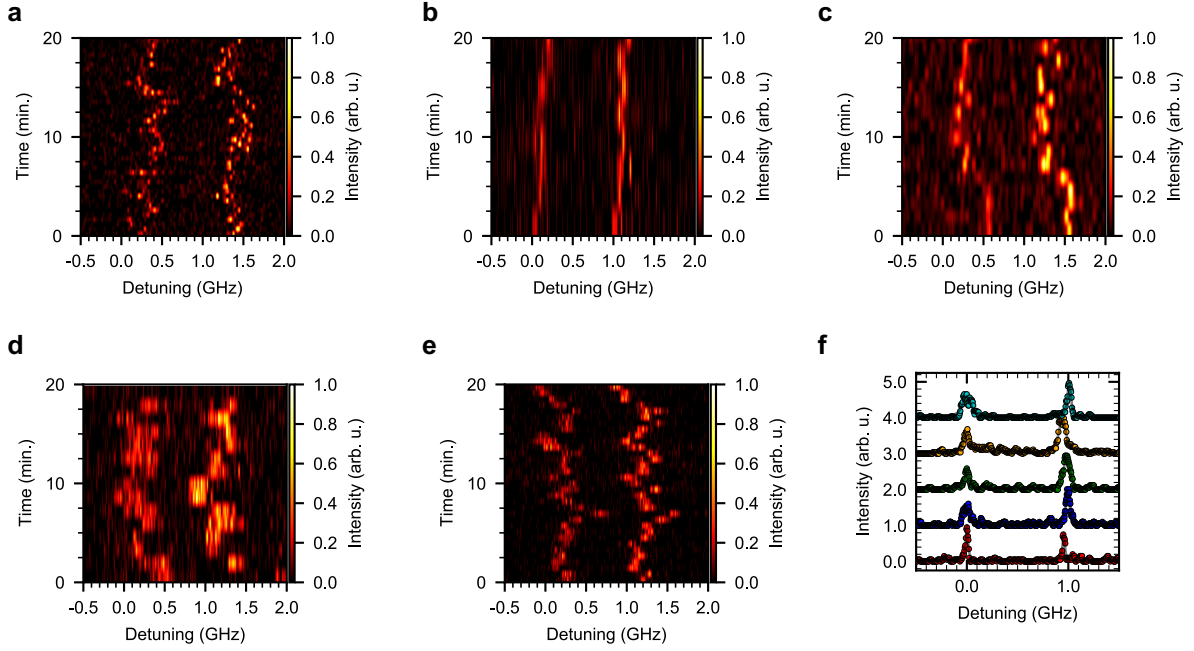

Figure S3. **a-e** Repeated resonant excitation scans during 20 minutes without repump laser for five implanted defects. **f** Single-line resonant excitation scans for five different V2 centres created via He$^+$ implantation. For better visibility, the A$_1$ optical transition for all spectra is centreed at zero detuning (the actual spectral distribution of the defects is $\pm 10$ GHz). Each measurement is additionally off-setted by one unit in $y$-direction. Linewidths are obtained via double-Lorentzian fits to the data. The obtained linewidths for the A$_2$ (A$_1$) transitions are (from bottom to top): $25 \pm 1$ MHz ($26 \pm 1$ MHz), $45 \pm 1$ MHz ($70 \pm 2$ MHz), $64 \pm 2$ MHz ($52 \pm 3$ MHz), $79 \pm 2$ MHz ($68 \pm 4$ MHz); $41 \pm 1$ MHz ($101 \pm 3$ MHz).

Remarkably, all defects show no signs of ionisation or blinking and single scans result in well-separated spin-selective lines. The very small remaining drift rate can be captured by our laser feedback system (see section S2 E). In Figure S3**f**, we display a single absorption line scans for the five additional defects from which we extract the linewidths of the A$_2$ and A$_1$ transitions via double-Lorentzian fit functions.

The average linewidth of the A$_2$ transition is $46 \pm 20$ MHz, where the uncertainty is taken from the standard deviation. Our results highlight the reproducibility of our implantation technique to generate high quality colour centres.

## B. Proton implantation

We investigate also V2 centre creation via proton implantation. As for the Helium implantation, we use a PMMA mask of 200 nm thickness on which 100 nm diameter holes are patterned using electron-beam lithography. Protons are implanted at an energy of 12 keV at a dose of $5 \cdot 10^{12} \, \text{cm}^{-2}$. The sample is subsequently annealed at 600°C in argon atmosphere for 30 minutes to remove some lattice damage and interstitial defects. A subsequent confocal microscopy image of the implanted defect array is shown in Figure S5a. Using the same approach as described in the section S1 A 1 for the Helium implanted defects, we infer the number of defects per spot by measuring the second-order correlation function. A typical $g^{(2)}$ signal is shown in Figure S5b. We collect statistical data for 110 implanted spots to infer the average number of V2 centres per spot (see Figure S5c). This translates to an implantation yield of $0.83 \pm 0.02\%$ for V2 centres. We also obtain the accuracy of the implantation, using the methods described in section S1 A 2. Similar to the above results, we find a lateral spatial variance of 56 nm, which is almost entirely dominated by the hole size (see Figure S5d). This is in accordance with SRIM simulations that predict a lateral projected straggle of 33 nm.

We subsequently measure the resonant absorption spectrum of five generated V2 defects. As shown in Figure S5a, single-line scans show spin-conserving transitions lines. From the double-Lorentzian fits, we extract the $A_2$ ($A_1$) linewidths, with four fits revealing nearly lifetime limited linewidths. Prolonged repeated measurements for 30 minutes, as seen in Figure S5b-f, show no signs of ionisation, but the drift rate is somewhat faster compared to the He$^+$ ion implanted defects. This is observed despite the fact that proton implanted defects are generated substantially deeper beneath the surface ($\sim 110$ nm for protons, compared to 40 nm for He$^+$). Thus, surface charge fluctuations are seemingly not the primary source of spectral drifts. Therefore, we attribute the increased drift rate for proton implanted defects to the increased relative formation of carbon versus silicon vacancies due to the use of lighter ions.

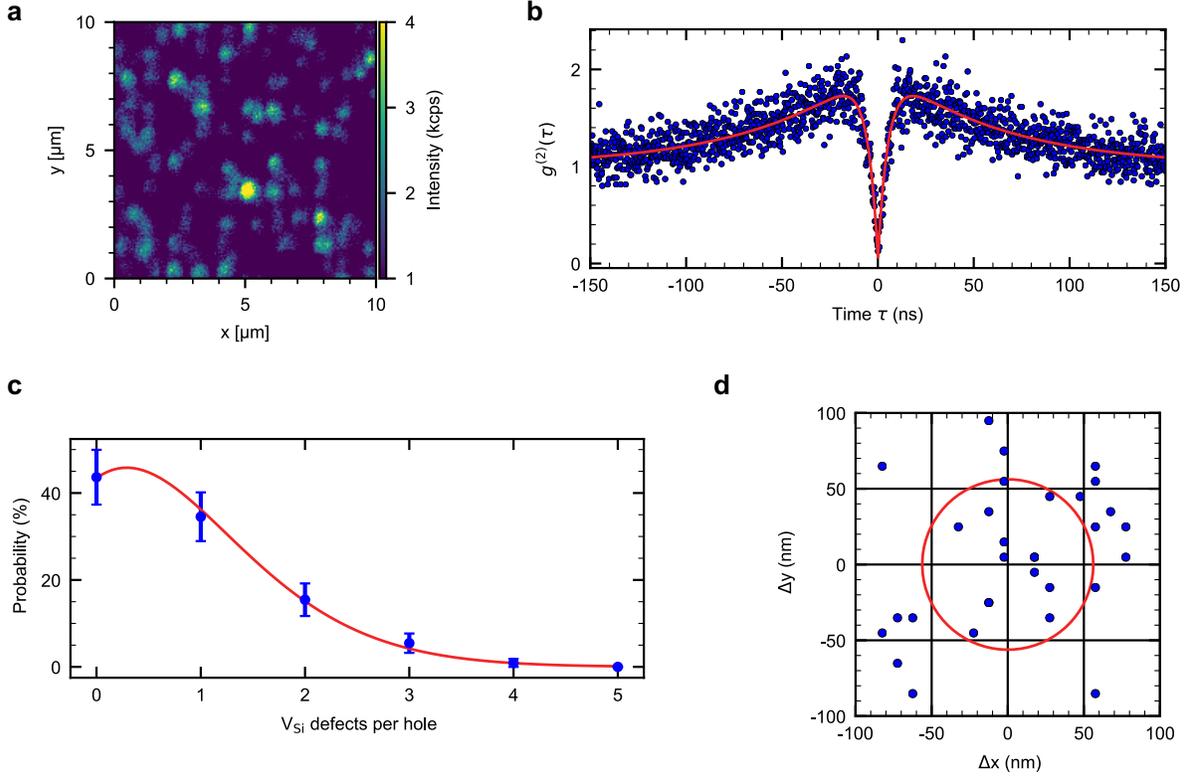

Figure S4. **a** Confocal fluorescence scan of the proton implanted defect centre array **b** Second order correlation function of a single V2 centre. From the fit to the data, we obtain a defect number of $N = 1.07 \pm 0.05$. **c** Probability distribution of the number of created V2 centres per site. The fit to the data is based on a Poissonian distribution. **d** Lateral spatial implantation accuracy compared to the simulated grid. The red circle represents the variance of 56 nm.

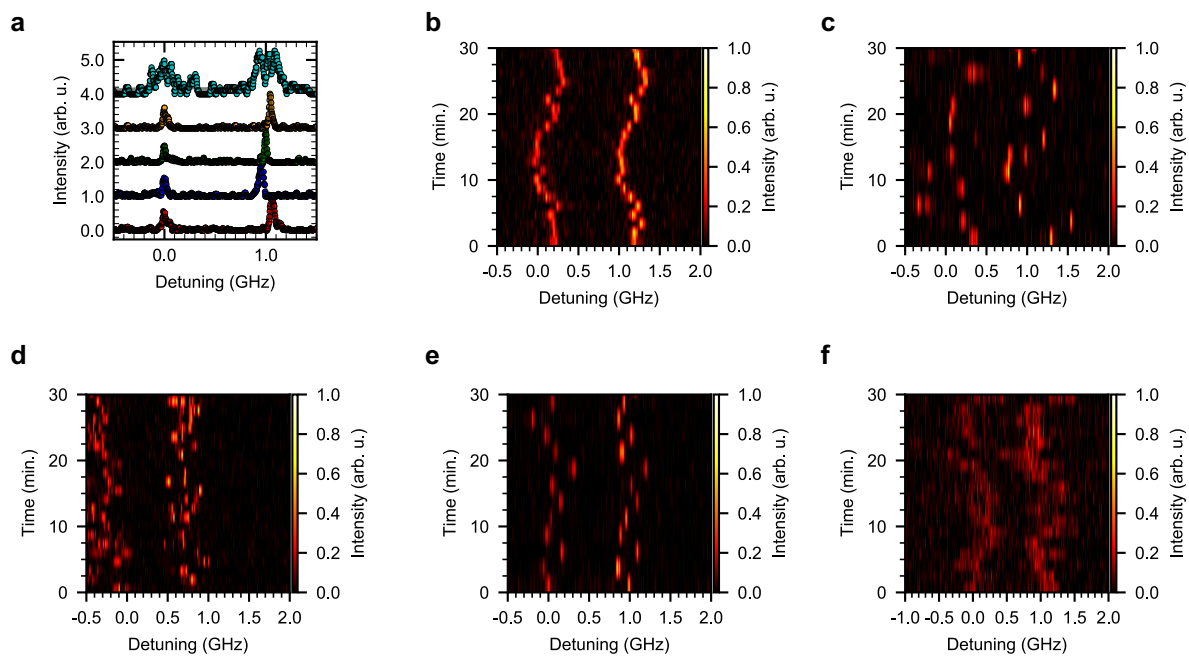

Figure S5. **a** Single-line resonant excitation scans for five different V2 centres created via proton implantation. For better visibility, the $A_1$ transition for all spectra is centreed at zero detuning. The actual spectral distribution of the defects is $\pm 10\,\mathrm{GHz}$. The single scan linewidths obtained from the fits are (from bottom to top): $43.9 \pm 0.6\,\mathrm{MHz}$ ($49 \pm 2\,\mathrm{MHz}$), $43 \pm 1\,\mathrm{MHz}$ ($39 \pm 1\,\mathrm{MHz}$), $20.9 \pm 0.4\,\mathrm{MHz}$ ($30 \pm 1\,\mathrm{MHz}$), $25.9 \pm 0.5\,\mathrm{MHz}$ ($37 \pm 1\,\mathrm{MHz}$); $226 \pm 6\,\mathrm{MHz}$ ($147 \pm 8\,\mathrm{MHz}$) for the $A_2$ ($A_1$) transition. **b-f** Repeated resonant excitation scans for 30 minutes without repump laser for five isolated V2 centres. No ionisation is observed. However, the drift rate is somewhat faster compared to the defects created via He$^+$ implantation.

## C. Silicon implantation

We implant Silicon ions ($Si^{2+}$) using a focussed ion beam device, operating at 35 keV acceleration voltage. The ion flux is varied from the minimal nominal dose of $1 \cdot 10^{-18}$ C per spot to $29 \cdot 10^{-18}$ C per spot, in other words, 3.1 to 91 ions are implanted per spot. After annealing the sample at 600°C for 30 minutes, we investigate the defect arrays at cryogenic temperatures ($T = 10$ K). Unfortunately, even at the lowest implantation dose, we did not observe any single defect.

We record photoluminescence emission spectra of the implanted spots using a Peltier cooled spectrometer (Ocean Insight QE pro, 0.3 nm resolution). A typical measurement is shown in Figure S6a. In all implanted spots, we clearly identify the characteristic zero-phonon lines of V1 and V2 centres a 862 nm and 917 nm, respectively. For V2 centres, we usually find ZPL emission linewidths of about 1 nm (see inset of Figure S6a), which is in contrast to the 0.3 nm emission linewidths that are obtained for $He^+$ and proton implanted defects. Thus, without surprise, we were not able to identify resonant absorption lines for those defects.

To further investigate the $Si^{2+}$ implanted defects, we perform additional measurements on the excited state lifetimes of the implanted defects. To this end, we use a 2-ps-pulsed excitation laser that operates at a central wavelength of 780 nm. From the time-dependent phonon sideband fluorescence, we extract the lifetime. The corresponding data is shown in Figure S6b. We clearly observe a bi-exponential decay with $1/e$-lifetimes of $\tau = 5.63 \pm 0.07$ ns and $\tau = 2.79 \pm 0.09$ ns, respectively. The longer decay constant closely resembles the typical results obtained for off-resonant excitation of V1 and V2 centres [S1, S3]. The shorter decay time may indicate that additional non-radiative decay channels have been created due to substantial crystal damage associated with heavy $Si^{2+}$ ion implantation [S4, S5]. Our hypothesis is strengthened by the single-decay results that are obtained with Helium and proton implantation (see main text). In future work, it may be interesting to reduce the Si ion flux to single silicon ions, and/or to see whether similar methods could be applied using helium ions.

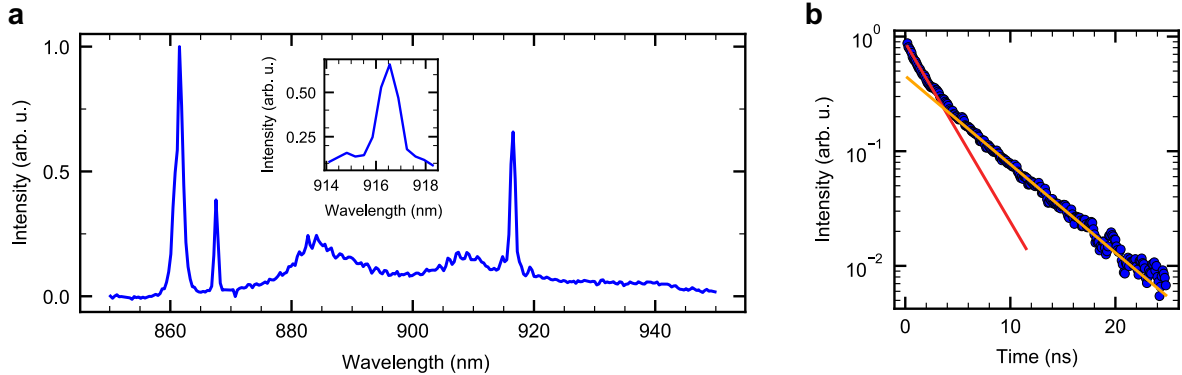

Figure S6. **a** Typical photoluminescence spectrum of an implantation spot using the lowest possible $Si^{2+}$ ion dose of $1 \cdot 10^{-18}$ C. The zero-phonon lines of both V1 and V2 are clearly visible, which clearly show that multiple defects are being generated. The inset shows a zoom-in for the V2 zero-phonon line, which is substantially wider compared to the proton and Helium implanted defects. **b** Pulsed off-resonant excitation lifetime measurement of a single implantation spot. The fit is based on a bi-exponential decay function, yielding $1/e$-lifetimes of $\tau = 2.79 \pm 0.09$ ns and $\tau = 5.63 \pm 0.07$ ns, respectively.

## S2. WAVEGUIDE INTEGRATED V2 CENTRES

### A. Defect density in the waveguide SiC sample

As mentioned in the Methods section, V2 centres have been created in the SiC sample via electron beam irradiation prior to waveguide fabrication. To estimate the expected defect density in the waveguides, we measure the defect density in the bulk material next to the waveguides via standard confocal microscopy. For the scans, we use a continuous-wave excitation laser with an operation wavelength of $\lambda = 785\,\text{nm}$. To estimate the defect density, we need to consider the axial resolution of the microscopy setup. Considering that our microscope objective has a numerical aperture of $NA = 0.9$, and that the refractive index of 4H-SiC is $n = 2.6$, we obtain an axial resolution of:

$$z_{\text{axial}} = \frac{2\,\lambda}{NA^2} \approx 1.9\,\mu\text{m}. \tag{S1}$$

We subsequently count 57 defects in a scanning area of $5 \times 15\,\mu\text{m}^2$ in the bulk next to the waveguides, see Figure S7. Assuming that the defects are axially homogeneously distributed over $1.9\,\mu\text{m}$ results in a defect density of $\rho_{\text{V}_{\text{Si}}} \approx \frac{57}{142.5\,\mu\text{m}^3} = 0.4\,\mu\text{m}^{-3}$.

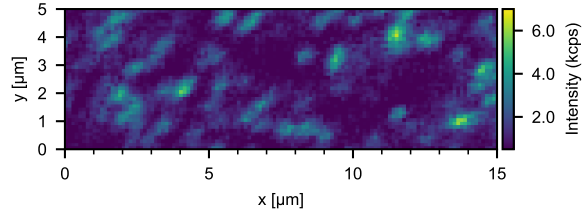

Figure S7. Confocal microscopy scan in the bulk next to the waveguides.

### B. Depth scans from waveguides to SiC surfaces

Considering the axial resolution of our microscopy setup, we underetch the waveguides by $\approx 2\,\mu\text{m}$ to ensure that optical excitation is performed on waveguide-integrated defects without spurious signal from defects in the SiC bulk substrate beneath. A typical confocal depth scan is shown in Figure S8a. Three waveguides are visible, with fluorescence of defects primarily occurring at the apices. The bulk SiC surface is also visible, including two near-surface defects. Fluorescent spots in waveguides and bulk SiC are well separated, thus confirming that resonant excitation experiments are indeed performed with defects localised in waveguides.

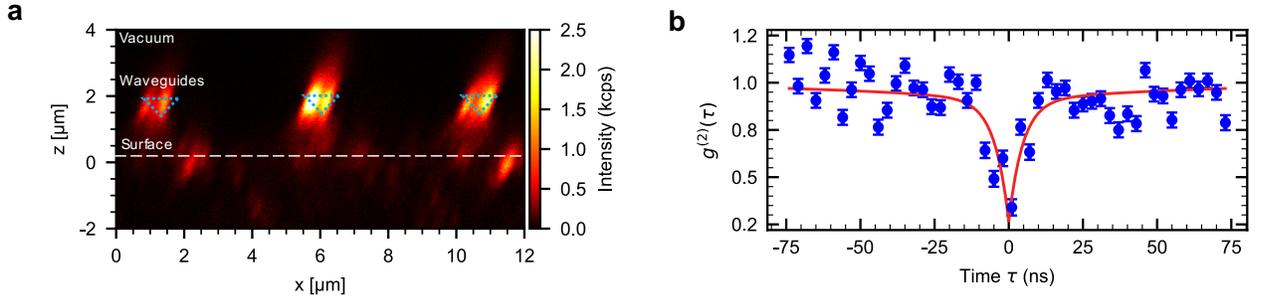

Figure S8. **a** Confocal microscopy depth scan in the waveguide area on the sample. The SiC bulk substrate surface is visible at a depth position $z \approx 0\,\mu\text{m}$. Three waveguides appear at $z \approx 2\,\mu\text{m}$. Defects in the bulk and waveguides appear well-separated, thus permitting selective excitation without spurious parasitic signals. **b** Second order correlation function of a single V2 centre in the waveguide, indicating single emitter behaviour ($g^{(2)}(\tau = 0) < 0.5$).

## C. Emission spectra of $V_{Si}$ and non-V2 colour centres in the waveguides

In Figure 2**f** in the main text, we show typical confocal scans along the nanofabricated waveguides at $T = 10\,\mathrm{K}$ with 785 nm continuous wave excitation. Multiple bright spots appear, however, most spots are not identified as V2 centres.

We have attempted to figure out the origin of the remaining fluorescent defects on the waveguides via measuring emission spectra. A typical emission spectrum of these defects and a typical spectrum of a V2 centre are shown in Figure S9. Unfortunately, no distinct features are present in the non-V2 centre spectra, despite the low temperature operation, making it hard for us to attribute the origin of the fluorescence to particular colour centres. At this point, we would like to mention that the identification of colour centre origin/structure is an ongoing field of research for the relatively new SiC platform. Additionally, many studies on unknown defects are performed at room temperature, which complicates a straightforward comparison to our low-temperature experiments. Nevertheless, comparing our spectra to known colour centres (e.g., as nicely summarized in reference [S6] and references therein), we can exclude carbon antisite vacancies (CAV), as well as annealing-related defects. Our measured emission spectra show some overlap with unknown defects that have been tentatively attributed to surface oxide [S7]. Such an oxide layer may form during the sample nanofabrication and/or acid cleaning. In the end, we would like to mention that the occasional presence of other colour centres on waveguides is not problematic for future experiments that target excitation of defects through the waveguides. Here, it is only important to be able to address individual VSi centres via spin-selective resonant optical excitation. The low powers used for resonant excitation experiments (nW range) are unlikely to induce any sizeable emission of other colour centres, even when performing excitation through the waveguide.

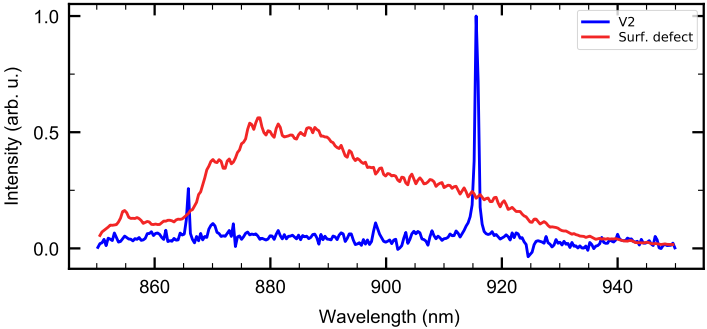

Figure S9. Typical emission spectra of colour centres in waveguides. The red curve represents a typical non-V2 defect, while the blue curve shows the emission spectrum of a V2 centre, which we identify in one out of five waveguides.

### D. PLE statistics

To underline the reproducibility of the results presented in the main text, we studied five additional V2 centres in waveguides. To this end, we show repeated resonant excitation scans for five additional defects in different waveguides. Experimental data is shown in Figure S10. All defects exhibit a very slow spectral drift and no ionisation is observed. The single-line scans of those five defects correspond to the upper five data sets in Figure 2**f** in the main text. The extracted single scan linewidths for the $A_2$ ($A_1$) transitions of those five defects are $21.4 \pm 0.3$ MHz ($49 \pm 2$ MHz), $123 \pm 7$ MHz ($115 \pm 20$ MHz), $17.6 \pm 0.3$ MHz ($32 \pm 1$ MHz), $78 \pm 2$ MHz ($216 \pm 8$ MHz), $15.6 \pm 0.4$ MHz ($44 \pm 3$ MHz), $93 \pm 1$ MHz ($100 \pm 5$ MHz).

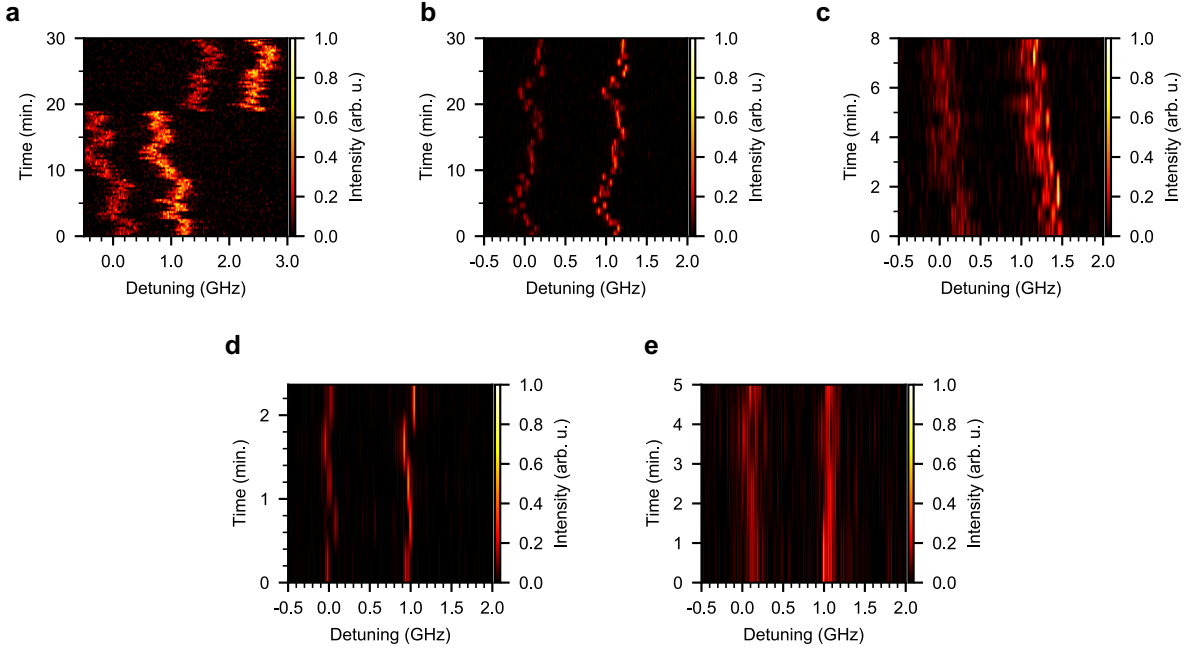

Figure S10. **a-e** Repeated resonant excitation scans for the defects presented in Figure 2**f** of the main text. No ionisation is observed and the remaining drift is assigned to surface charge fluctuations.

## E. Laser stabilisation protocol

The resonant absorption lines of our nanofabricated V2 centres still exhibit a very slow drift, e.g., as shown in Figure S10. To perform complex long-term measurements, we thus periodically refocus the resonant excitation laser on the $A_2$ optical transition. The refocussing period is typically chosen on a time scale of a few to several minutes. The employed protocol is schematically depicted in Figure S11. During all steps in the protocol, we provide a microwave (MW) drive resonant with the ground state transitions $m_S = +1/2 \leftrightarrow m_S = +3/2$ and $m_S = -1/2 \leftrightarrow m_S = -3/2$ to counteract spin pumping by mixing the ground state populations. The laser feedback protocol goes as follows:

(1) We probe whether the absorption lines drifted by applying a resonant excitation on the $A_2$ optical transition for $500\,\mathrm{ms}$. Provided that more than 300 photons are detected, we consider that no drift occurred, and the main measurement is continued. If less than 300 photons are detected, we proceed with step (2).

(2) We perform up to two resonant excitation scans (PLE), by tuning the resonant excitation laser over a frequency range of 3 GHz. The resulting fluorescence signal is fitted with a double-Lorentzian function. The goodness of the fit is evaluated by several criteria, such as the peak separation that must be within $0.9 - 1.1$ GHz, the peak ratio that must be in the range of $\frac{1}{3}$ to 3, the peak width that must be in the range of 10 to 100 MHz, and the signal-to-background ratio that must exceed 5. If all criteria are met, we fix the laser on the high-frequency transition ($A_2$ line) and resume the main measurement. The laser scanning window is also reset such that the $A_2$ transition occurs at $2/3$ of the (potential) next scan. If the above criteria are not met, we repeat step (2) one more time. If the criteria are not met after the second attempt, we proceed to step (3).

(3) An off-resonant laser pulse ($1\,\mathrm{mW}$, $1\,\mathrm{s}$, $785\,\mathrm{nm}$) is applied to modify/reset the charge environment of the V2 centre. Then, we perform up to two resonant excitation scans with a frequency range of 13 GHz. If the signal can be fitted by a double-Lorentzian function with the same fit criteria as mentioned in (2), we continue with step (2). Note that we do not resume the main measurement directly due to laser drift and hysteresis after extended scan ranges. If step (3) is not successful after two attempts, we proceed with step (4).

(4) Similar to step (3), we step (4) employs the same off-resonant excitation pulse, however followed by a resonant excitation scan covering 20 GHz. This scan range is sufficient to always identify the absorption lines, e.g., we have never observed spectral wandering and distribution beyond this range [S8]. Once step (4) is successful, we proceed to step (2).

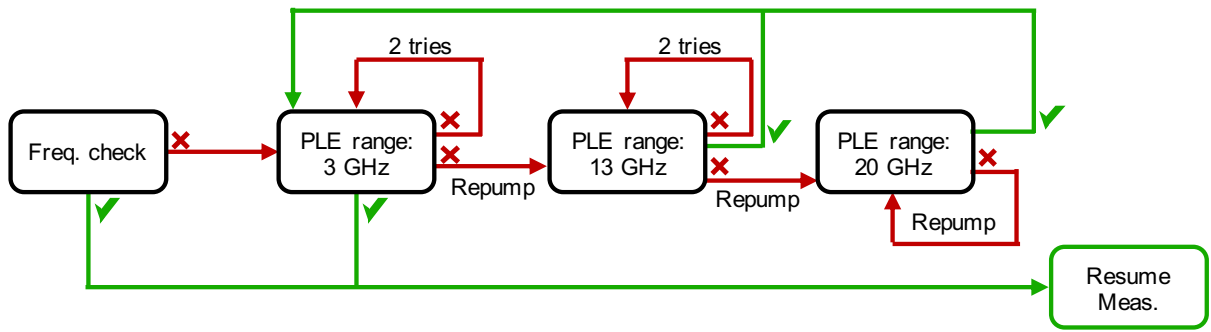

Figure S11. Schematic of the protocol to periodically refocus the resonant laser on the $A_2$ optical transition of the waveguide-integrated V2 centres. The green (red) arrows correspond to successful (unsuccessful) steps in finding the $A_2$ transition.

## F. Spin properties in bulk material

To infer the spin coherence degradation that V2 centres experience after waveguide fabrication, we compare their spin properties to deep bulk defects that have not been affected by the nanofabrication processes. In the following, we perform Hahn-echo and Ramsey interferometry measurements on a V2 centre that resides in the bulk in a non-etched area on the sample. The results are shown in Figure S12**a-b**. From the fit to the data, we obtain a dephasing time of $T_{2,bulk}^* = 21 \pm 1$ $\mu$s and a Hahn echo coherence time of $T_{2,bulk} = 0.84 \pm 0.03$ ms. Those times are very similar to the ones measured for the waveguide integrated V2 centres, which corroborates that our nanofabrication recipes induce minimal damage.

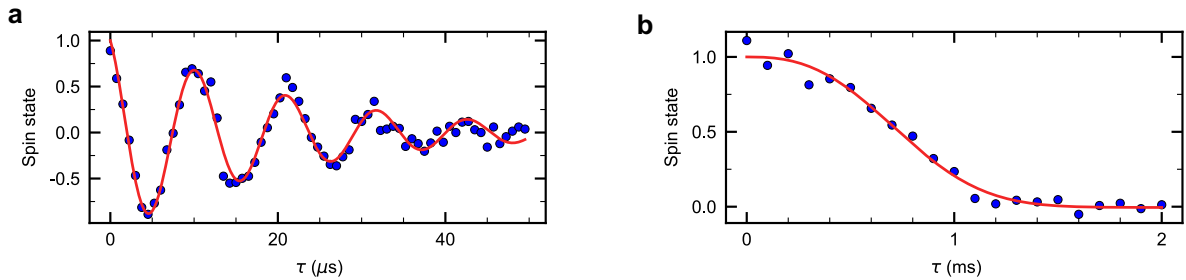

Figure S12. **a** Spin dephasing time for V2 centres in the bulk. From the fit to the data, we obtain $T_{2,bulk}^* = 21 \pm 1$ $\mu$s. **b** Spin coherence time of bulk V2 centres using the Hahn echo sequence. The fit to the data results in a coherence time of $T_{2,bulk} = 0.85 \pm 0.03$ ms.

## G. Optimized waveguide design considering the demonstrated implantation techniques

Our implantation experiments demonstrated that the excellent spin-optical properties of V2 centres are not degraded by He$^+$ ion implantation and when integrating the defects in nanophotonic waveguides. This promises the development of highly efficient and robust spin-photon interfaces via system integration into waveguides, photonic crystal cavities, and quantum photonic circuits in general. In this section, we discuss ideal waveguide designs based on the demonstrated implantation and nanofabrication techniques.

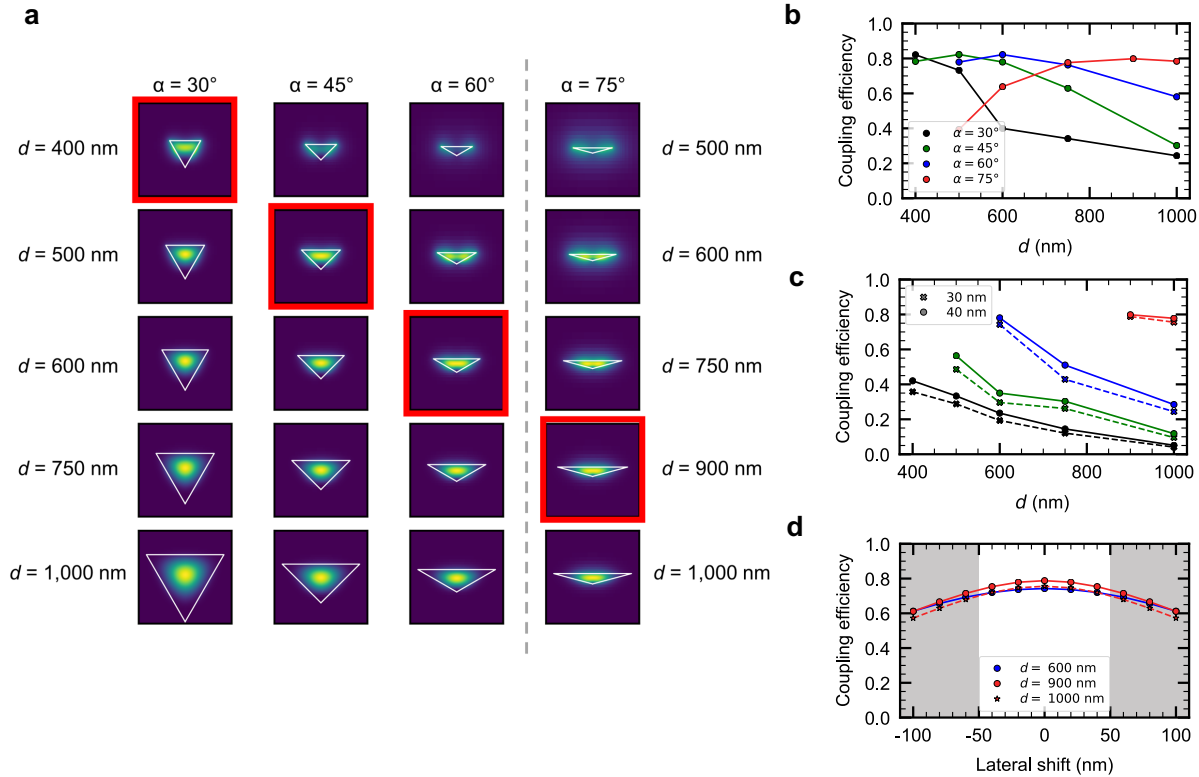

Figure S13. Modelled colour centre emission coupling into a triangular cross section waveguide. **a** TE mode profiles of triangular waveguides with variable etch angle $\alpha$ and width $d$. Highest waveguide-coupling efficiency is achieved for the devices that are highlighted by a red box. **b** The coupling efficiency of light emitted by a centrally located V2 centre for variable waveguide geometry. **c** The coupling efficiency of light emitted by a V2 centre on the vertical symmetry axis for variable emitter depth and waveguide geometry; the colour legend in **b** applies to the plot. **d** The coupling efficiency of light emitted by a 30 nm deep V2 centre laterally shifted from the central position for variable waveguide geometry; the colour legend in **b** applies to the plot.

In colour centre integration with triangular waveguides, highly efficient single-mode light propagation is a crucial prerequisite for quantum applications. The relationship between device profile and its supported mode wavelength has been modeled recently [S9]. Here, we expand on those findings to propose efficient design guidelines for V2 centre emission propagation, robust to the demonstrated implantation uncertainties. In particular, we analyze the coupling efficiency of light emitted by a centrally located horizontal dipole at 917 nm into the transverse electric (TE) modes of a triangular waveguide. We find the existence of the preferential waveguide width $d$ for each studied etch angle $\alpha$, as presented in Figure S13**a,b**. Interestingly, the mode shape of the optimal device appears more rectangular than circular, and localized within silicon carbide. For the waveguides fabricated in this work ($\alpha = 45°$, $d = 1000\,\text{nm}$) the coupling efficiency into the fundamental TE mode is approximately 30% (15% in each propagation direction in the waveguide). The efficiency would rise to 82% if those waveguides were minimized to 500 nm width. We note that these coupling efficiencies assume a defect in the centre of the waveguide, i.e., at a depth of 125 nm along the vertical symmetry axis of the devices with $\alpha = 45°$ and $d = 500\,\text{nm}$. Figure S13**a,b** show additional results for V2 centres positioned in at the centre of different waveguide geometries ($\alpha = 30° - 75°$, $d = 400\,\text{nm} - 1000\,\text{nm}$). Devices with optimal collection efficiency are highlighted by a red box in Figure S13**a**. Figure S13**b** shows further that waveguides with steeper etching angle $\alpha$ provide high optimal efficiencies over a larger range in $d$.

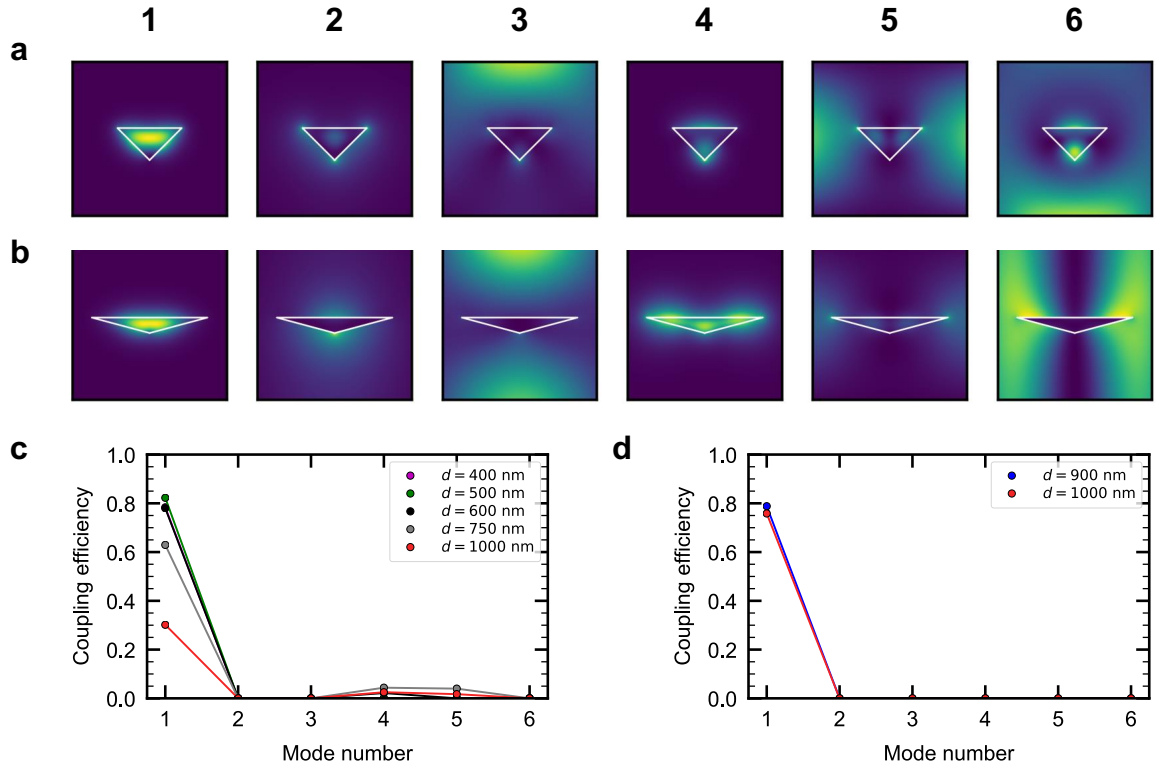

Figure S14. Single mode operation of optimized triangular waveguides. **a** Profiles of the six lowest energy modes supported in a triangular waveguide with etch angle $\alpha = 45°$ and width $d = 500\,\text{nm}$ and **b** etch angle $\alpha = 75°$ and width $d = 900\,\text{nm}$; mode #1 is the fundamental TE polarized mode. **c** The coupling efficiency of light emitted by a centrally located horizontally polarized dipole into each of the modes shown under **a**; the coupling efficiency is evaluated after $10\,\mu\text{m}$ propagation. Single mode operation is observed for waveguide widths of 400 nm and 500 nm. **d** The coupling efficiency of light emitted by a 30 nm deep horizontally polarized emitter to each of the modes shown under b showing single mode operation.

To account for the demonstrated He$^+$ ions implantation technique that results in a preferred V2 centre creation at $30 - 40\,\text{nm}$ depth, we now provide additional simulation results to identify the optimal geometry. As presented in Figure S13**c**, the coupling efficiency does not change significantly in the investigated depth range ($d = 30 - 40\,\text{nm}$). Devices with higher etch angles result in consistently better coupling efficiencies (up to 80%). Remarkably, the $\alpha = 75°$ devices show very robust coupling efficiencies, especially with regards to the waveguide width $d = 900 - 1000\,\text{nm}$, as shown in Figure S13**c**. We also consider the demonstrated lateral defect creation accuracy which was determined to be $\pm 54\,\text{nm}$ in this work and mainly limited by the implantation mask's hole size (100 nm diameter). As shown in Figure S13**d**, lateral displacement of the emitters does not play a significant role, i.e., fundamental TE mode coupling efficiencies remain within 10% of their peak value. The lost photons couple to the fundamental transverse magnetic (TM) mode and can, in practice, be filtered out by polarisation.

For optimal coupling of light out of the waveguides into single-mode optical fibres, it is important to ensure single TE mode operation. To this end, we simulate the first six TE modes for devices with ($\alpha = 45°$, $d = 500\,\text{nm}$) and ($\alpha = 75°$, $d = 900\,\text{nm}$), respectively. The results are shown in Figure S14**a** and **b**, respectively. The coupling efficiency of from a 30 nm deep emitter to the different modes and varying $d$ is shown in Figure S14**c** and **d**, respectively. We find that both preferred geometries ($\alpha = 45°$, $d = 500\,\text{nm}$) and ($\alpha = 75°$, $d = 900\,\text{nm}$), provide $\approx 80\%$ coupling efficiency into the first TE mode without noticeable excitation of higher-order modes.

Our modelling results show that the demonstrated Faraday cage etching techniques and He$^+$ ion implantation provide a rich toolset for integrated SiC colour centre photonics. The single TE mode operation of triangular waveguides is highly efficient and robust in multiple degrees of freedom: vertical and lateral emitter displacement, fabrication dimension and surface roughness imperfections. Moreover, these waveguides can be coupled to triangular photonic crystal cavities which are known to provide strong Purcell enhancement utilized to boost the generation of indistinguishable single photons in colour centres [S9–S11].

### S3. PHOTONIC COUPLING IN THE WAVEGUIDE AND OBSERVATION OF LESS V2 CENTRES INSIDE THE WAVEGUIDES

Finite-Difference Time-Domain (FDTD) method is particularly suitable for simulating the expected performance of nanophotonic devices, processes, and materials. This method models the propagation of light by discretizing Maxwell's equations in both time and space coordinates in a leap-frog manner. Owing to high accuracy and broad-frequency response, FDTD method has been widely used as a verification tool for photonic devices over the years. Moreover, FDTD method has been employed to simulate nanocavities in triangular diamond waveguides [S11, S12]. Recently, we have used FDTD modeling to analyze color center positioning and Purcell enhancement in triangular SiC waveguides and nanobeam cavities [S9]. In this work, as well as in the present manuscript, we have used Lumerical software to perform 3D FDTD electromagnetic simulations, an accurate and widely used tool in academia and industry.

In an attempt to infer the photonic coupling efficiency along the waveguide using our present optical arrangement (collection objective from the top), we performed additional experimental and simulation studies. The basic idea was to infer saturation count rates from V2 centres in the bulk and compare these count rates to emitters in waveguides. Assuming that the total photon emission rate in both cases does not change, and comparing with results from simulations, would then allow to confirm photonic coupling into the waveguide.

Our FDTD model assumes a horizontal dipole emitter in a SiC triangular waveguide (or bulk) studies the light propagation equivalent to the experimental setup. The collection of light into an objective of NA = 0.9 is calculated in a two-step analysis. In the first step, the fraction of light that leaves silicon carbide upwards is evaluated. In the second step, Fourier transform of the electromagnetic field above the structure produces the far-field pattern in k-space, and can be used to calculate the fraction of light that enters the NA = 0.9 objective. The modelling results support the experimental findings and are detailed in Figure S15.

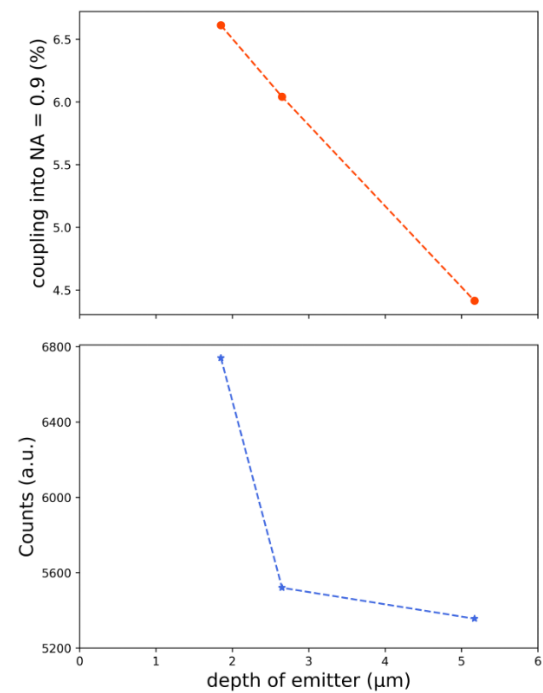

Figure S15. Model of an emitter in bulk. The model of a silicon vacancy in SiC bulk at arbitrary depth (top) shows a decreasing trend in collected photons with the increasing depth of the emitter. Experimental results (below) show a similar trend.

In the next step, we performed simulations on the out-coupling efficiency for defects located in 45° underetched 1000 nm wide waveguides. We found that the collectable photon rate depends very strongly on the exact location of the emitter in the waveguides, see Figure S16.

Considering the large variability of achievable collection efficiency and the limited accuracy towards inferring each defect's precise position inside the waveguide, made it impossible to faithfully measure the saturation count rate, which was therefore not attempted.

These simulations provide also a plausible explanation for the observation of less V2 centres in the waveguides compared to the defect density in the bulk. In this sense, we note that we only identify V2 centres that are bright compared to the waveguide's background fluorescence. This means that we observe only these defects that are placed at ideal positions in the waveguide, notably those that are close to the waveguides' apices, see Figure S16. It is thus very likely that the actual density of negatively charged V2 centres is not modified significantly in the waveguides. As the observed V2 centres show stable optical lines and no ionization, this could indicate that bandbending near the surfaces and edges is not significant.

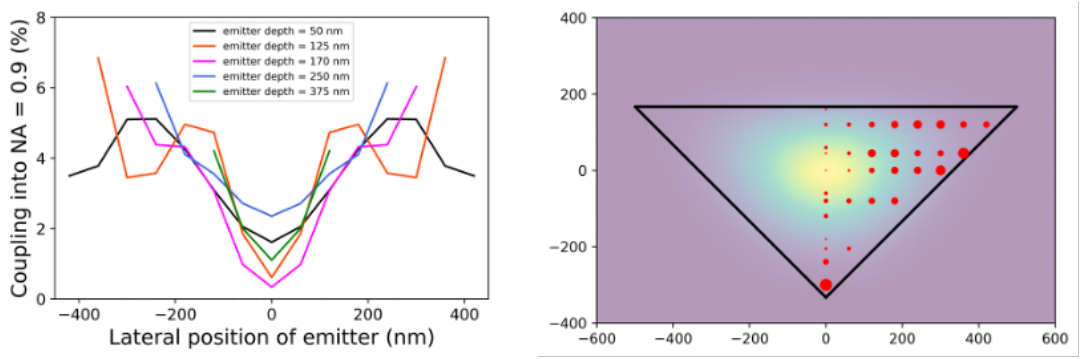

Figure S16. Model of an emitter in a triangular waveguide. The model of a silicon vacancy in a triangular SiC waveguide reveals that collection efficiency into an objective is higher at the edges of the structure than in its center (left). This data supports the experimental findings where emitters are identified at the sides of the structure (see Figure 2f in the Main Text, as well as Figure S8). A plausible explanation is that the emitters that couple to the fundamental mode do not scatter light toward the objective (right). The red dot size is proportional to the light collected by the objective for emitter at that location.

## S4. ELECTRON-NUCLEAR SPIN DYNAMICS

In this section we develop a framework to describe the spin dynamics of a weakly coupled electron-nuclear spin pair during the Hahn echo and CPMG sequences. Further, we provide a methodology to estimate the nuclear spin gate fidelities. We apply the techniques to our hybrid system comprising an electron spin-$\frac{3}{2}$ system coupled to a nuclear spin-$\frac{1}{2}$. However, our framework can be straightforwardly applied to arbitrary electron-nuclear spin systems.

### A. Decoupling sequences

#### 1. Hamiltonian

We consider a weakly coupled electron nuclear spin system, such that the secular approximation is valid. Therefore, the Hamiltonian can be expressed as:

$$\hat{\mathcal{H}} = \sum |s_i\rangle \langle s_i| \otimes \omega_i \mathbf{n}_i \cdot \hat{\mathbf{I}} = \sum |s_i\rangle \langle s_i| \otimes \hat{\mathcal{H}}_i. \tag{S2}$$

Here, $|s_i\rangle$ denote the different electron spin levels, and $\hat{\mathbf{I}}$ is the spin matrix representation of the nuclear spin. $\omega_i$ and $\mathbf{n}_i$ are the angular frequency and the rotation axis of the nuclear spin when the electron spin is in the state $|s_i\rangle$, respectively. The terms can be expressed as a function of the hyperfine coupling terms as:

$$\omega_i = \sqrt{(s_i A_\parallel - \gamma_n B)^2 + (s_i A_\perp)^2}, \tag{S3}$$

$$\mathbf{n}_i = \frac{1}{\omega_i}(s_i A_\perp, 0, s_i A_\parallel - \gamma_n B). \tag{S4}$$

Here, $A_\parallel$ and $A_\perp$ are the parallel and perpendicular hyperfine coupling terms, respectively. $\gamma_n$ is the nuclear spin gyromagnetic moment, and $B$ the external magnetic field strength (assumed here to be parallel to the $z$-axis, which is the electron spin quantisation axis). With this, we now develop the nuclear spin evolution operator with the electron spin state being in the state $|s_i\rangle$ ($\hbar = 1$):

$$\hat{\mathcal{U}}_i(\tau) = \exp\left[-\mathrm{i}(\omega_i \tau)\mathbf{n}_i \cdot \hat{\mathbf{I}}\right]. \tag{S5}$$

In the Bloch sphere representation, this corresponds to a rotation by an angle $\omega_i \tau$ around the axis $\mathbf{n}_i$.

#### 2. Dynamics

The hyperfine coupling can be used to drive the nuclear spin. For a weakly coupled nuclear spin, so-called dynamical decoupling sequences provide a powerful means. By alternating the electron spin between the two states $|s_0\rangle$ or $|s_1\rangle$ (in our case $|s_0\rangle = |m_\mathrm{S} = +1/2\rangle$ and $|s_1\rangle = |m_\mathrm{S} = +3/2\rangle$), the nuclear spin experiences different rotations. We consider the sequence in which the electron spin is initialized into the state $|\psi\rangle_\mathrm{init} = |s_0\rangle$, followed by a $\pi/2$-pulse to obtain the superposition state $(|s_0\rangle + |s_1\rangle)/\sqrt{2}$. Thereafter, we apply the spin manipulation sequence $(\tau - \pi - \tau)^N$, in which, $\tau$ represents a waiting time, $\pi$ a spin-flip pulse, and $N$ is an integer number. If we denote the initial nuclear spin state as $|n\rangle$, then the bipartite system's state at the end of the sequence is:

$$|\Psi\rangle = \frac{1}{\sqrt{2}}\left(|s_0\rangle \otimes (\hat{U}|n\rangle) + |s_1\rangle \otimes (\hat{V}|n\rangle)\right). \tag{S6}$$

Here, $\hat{U}$ ($\hat{V}$) describes the nuclear spin evolution with the electron spin state being $|s_0\rangle$ ($|s_1\rangle$). To simplify the reading, we assume an additional $\pi$-pulse at the end of the sequence for even $N$. This way, we preserve the electron spin state in the case of no nuclear spin being present. After a final $\pi/2$-pulse, the initial electron-spin state $|\psi\rangle_\mathrm{init} = |s_0\rangle$ is preserved with a probability of:

$$P(s_0) = \frac{1 + \cos\left(\frac{\theta_{\hat{V}^\dagger\hat{U}}}{2}\right)}{2}.$$

(S7)

Here, $\theta_{\hat{V}^\dagger\hat{U}}$ is the rotation angle of the nuclear spin during the operation $\hat{V}^\dagger\hat{U}$.

**B.   Hahn echo and modulation**

The Hahn echo is a particular case of the decoupling sequence with $N = 1$, in which $\hat{U}$ and $\hat{V}$ are simply:

$$\hat{U} = \hat{\mathcal{U}}_1 \cdot \hat{\mathcal{U}}_0. \tag{S8}$$

$$\hat{V} = \hat{\mathcal{U}}_0 \cdot \hat{\mathcal{U}}_1. \tag{S9}$$

The operators $\hat{\mathcal{U}}_i$ are described in equation (S5). Taking advantage of trigonometric identities

$$\cos(\theta_{1,0}/2) = \cos(\theta_1/2)\cos(\theta_0/2) - \sin(\theta_1/2)\sin(\theta_0/2)\mathbf{n}_1 \cdot \mathbf{n}_0. \tag{S10}$$

$$\sin(\theta_{1,0}/2)\mathbf{n_{1,0}} = \cos(\theta_1/2)\sin(\theta_0/2)\mathbf{n}_0 + \cos(\theta_0/2)\sin(\theta_1/2)\mathbf{n}_1 + \sin(\theta_1/2)\sin(\theta_0/2)\mathbf{n}_1 \times \mathbf{n}_0, \tag{S11}$$

we obtain the well-known equation:

$$P(s_0) = 1 - \frac{k}{4}\left(2 - 2\cos(\omega_0\tau) - 2\cos(\omega_1\tau) + \cos(\omega_+\tau) + \cos(\omega_-\tau)\right). \tag{S12}$$

Here, $\omega_\pm = \omega_0 \pm \omega_1$, and $k$ being the so-called modulation depth parameter:

$$k = \left(\frac{\omega_L A_\perp}{\omega_0 \omega_1}\right)^2, \tag{S13}$$

with $\omega_L = \gamma_n B$ being the nuclear spin Larmor frequency.

## C. CPMG resonances

For an even number $N$ of $\pi$-pulses, the decoupling sequence can be written as: $(\tau\text{-}\pi\text{-}\tau\text{-}\tau\text{-}\pi\text{-}\tau)^{N/2}$, and the operators $\hat{U}$ and $\hat{V}$ as:

$$\hat{U} = \left(\hat{\mathcal{U}}_0 \cdot \hat{\mathcal{U}}_1 \cdot \hat{\mathcal{U}}_1 \cdot \hat{\mathcal{U}}_0\right)^{N/2} = \left(\hat{\mathcal{U}}_{D,1} \cdot \hat{\mathcal{U}}_{D,0}\right)^{N/2}, \tag{S14}$$

$$\hat{V} = \left(\hat{\mathcal{U}}_1 \cdot \hat{\mathcal{U}}_0 \cdot \hat{\mathcal{U}}_0 \cdot \hat{\mathcal{U}}_1\right)^{N/2} = \left(\hat{\mathcal{U}}_{D,0} \cdot \hat{\mathcal{U}}_{D,1}\right)^{N/2}. \tag{S15}$$

Here, $\hat{\mathcal{U}}_{D,i}$ is the composed rotation operator applied to the nuclear spin with the electron spin being in the state $|s_i\rangle$ for a time $\tau$ and additionally in the state $|s_{1-i}\rangle$ for another time $\tau$.

To optimize nuclear spin manipulation fidelity, we are now interested in the optimum (resonant) waiting time $\tau_r$ at which $\hat{U}$ and $\hat{V}$ are anti-parallel. From equation (S11), we obtain anti-parallel rotation axes if, and only if:

$$\cos(\theta_{D,0}) = \cos(\theta_{D,1}) = 0 \qquad \& \qquad \mathbf{n}_{D,0} \times \mathbf{n}_{D,1} \neq 0. \tag{S16}$$

With equation (S10), we obtain the condition for anti-parallel rotation axes:

$$\tan(\omega_0 \tau_r/2) \tan(\omega_1 \tau_r/2) = (\mathbf{n}_0 \cdot \mathbf{n}_1)^{-1} \tag{S17}$$

Interestingly, for periodic waiting times $\tau_k = \frac{(2k-1)\pi}{(\omega_0+\omega_1)}$ with $k \in \mathbb{N}^+$, we have $\tan(\omega_0\tau_k/2)\tan(\omega_1\tau_k/2) = 1$. We find further that in the case of negligible perpendicular hyperfine coupling ($|A_\perp| \ll |A_\parallel|, |\omega_L|$), and with the definition of $\mathbf{n}_i$ in equation (S4), both rotation axes are along the $z$-axis. The inverse vector product is then $(\mathbf{n}_0 \cdot \mathbf{n}_1)^{-1} \approx 1$. Therefore, we find:

$$\tau_r = \tau_k = \frac{(2k-1)\pi}{\omega_0 + \omega_1}. \tag{S18}$$

This equation represents the zero order approximation, which is exact in the asymptotic cases $B \to \infty$ or $A_\perp = 0$. For more realistic and general cases, the equation needs to be extended with correction terms. In the next section we develop the analytic expression for the first order correction term in $(A_\perp/B)^2$, and we define the criteria for the validity range of the approximation.

### 1. 1st order correction

We assume $\tau = \tau_k(1 + \epsilon_\tau)$, with $\epsilon_\tau \ll 1$. The left side in equation (S17) can be written as a first-order Taylor expansion:

$$\tan(\omega_0 \tau/2)\tan(\omega_1 \tau/2) = 1 + \epsilon_\tau \cdot \frac{(2k-1)\pi}{\sin(\omega_0 \tau_k/2)} + \mathcal{O}(\epsilon_\tau^2). \tag{S19}$$

Then, solving equation (S17) for $\epsilon_\tau$ results in:

$$\epsilon_\tau = \sin(\omega_0 \tau_k/2)\frac{(\mathbf{n}_0 \cdot \mathbf{n}_1)^{-1} - 1}{(2k-1)\pi}. \tag{S20}$$

Considering the vector product in the numerator, one sees from this equation that $\epsilon_\tau \to 0$ for $B \to \infty$. We now develop an expression to define the magnetic field range $B_{\text{crit}} \leq B \leq \infty$ in which the approximation is valid. In other words, given a nuclear spin with coupling terms $A_\parallel$ and $A_\perp$, which is the minimum required magnetic field $B_{\text{crit}}$ in which the approximation $\tau_r = \tau_k(1 + \epsilon_\tau)$ is valid? To have a negligible contribution from all higher order terms in $\mathcal{O}(\epsilon_\tau^2)$, we want $\epsilon_\tau^2 \ll \epsilon_\tau$, which translates to $(\mathbf{n}_0 \cdot \mathbf{n}_1)^{-1} - 1 \ll \pi$.

### 2. Validity range for the approximated resonance approximation

From equation (S4), we have:

$$(\mathbf{n}_0 \cdot \mathbf{n}_1)^{-1} = \sqrt{1 + \left( \frac{A_\perp \cdot \omega_L}{s_1 s_0 \cdot (A_\perp^2 + A_\parallel^2) - (s_0 + s_1) \cdot A_\parallel \cdot \omega_L + \omega_L^2} \right)^2}. \tag{S21}$$

Therefore, we can define an equivalence as:

$$(\mathbf{n}_0 \cdot \mathbf{n}_1)^{-1} - 1 \leq \epsilon_n \qquad \Leftrightarrow \qquad \left| \frac{A_\perp \cdot \omega_L}{s_0 s_1 \cdot (A_\perp^2 + A_\parallel^2) - (s_0 + s_1) \cdot A_\parallel \cdot \omega_L + \omega_L^2} \right| \leq \sqrt{(1 + \epsilon_n)^2 - 1}. \tag{S22}$$

Here, $\epsilon_n \ll 1$ is a parameter that is related to the magnitude of the correction term in equation (S19). The last inequality is solved for two solutions:

$$\omega_L + A_- + s_0 s_1 \frac{A_\perp^2 + A_\parallel^2}{\omega_L} \geq 0, \text{ and} \tag{S23}$$

$$\omega_L + A_+ + s_0 s_1 \frac{A_\perp^2 + A_\parallel^2}{\omega_L} \leq 0, \tag{S24}$$

with $A_\pm = (s_0 + s_1)A_\parallel \pm \frac{|A_\perp|}{\sqrt{(1+\epsilon_n)^2-1}}$. As required above, we are interested in the solution that covers the case $B \to \infty$, ($B_{\mathrm{crit}} \leq B \leq \infty$). Because $\omega_L = \gamma_n B$, the correct set of solutions depends on the sign of the gyromagnetic ratio $\gamma_n$. If $\gamma_n$ is positive (negative), the inequality to solve is equation (S23) (equation (S24)). To determine the critical magnetic field $B_{\mathrm{crit}}$, we use the following equation:

$$\omega_L^2 + A_{\mathrm{sign}(\gamma_n)} \cdot \omega_L + s_0 s_1 (A_\perp^2 + A_\parallel^2) = 0, \tag{S25}$$

in which $\mathrm{sign}(\gamma_n) = \gamma_n/|\gamma_n|$. Finally we have:

$$(\mathbf{n}_0 \cdot \mathbf{n}_1)^{-1} - 1 \leq \epsilon_n \qquad \Leftrightarrow \qquad B \geq B_{\mathrm{crit}} = \frac{A_{\mathrm{sign}(\gamma_n)} + \mathrm{sign}(\gamma_n)\sqrt{A_{\mathrm{sign}(\gamma_n)}^2 - 4 s_0 s_1 (A_\perp^2 + A_\parallel^2)}}{2\gamma_n}. \tag{S26}$$

Here, we have:

$$A_{\mathrm{sign}(\gamma_n)} = (s_0 + s_1)A_\parallel + \mathrm{sign}(\gamma_n) \frac{|A_\perp|}{\sqrt{(1 + \epsilon_n)^2 - 1}}. \tag{S27}$$

We now limit the first-order correction to 10% ($\epsilon_\tau \leq 0.1$) to keep higher order terms at a negligible level. This means that $\epsilon_n = 0.1 \cdot \pi$. With this, we can express the denominator in equation (S27) as $1/\sqrt{(1 + \epsilon_n)^2 - 1} \approx 1.17$.

We thus obtain the following theorem for the first-order approximation of the CPMG resonance times $\tau_r$:

**Theorem.** *For given nuclear spin hyperfine coupling terms $A_\parallel$ and $A_\perp$, and in an external magnetic field along the z-axis with strength $B$, the CPMG resonances occur at:*

$$\tau_{\mathrm{approx}}^{(k)} = \frac{(2k-1)\pi}{\omega_0 + \omega_1} \left( 1 + \sin\left( \frac{(2k-1)\pi\omega_0}{2(\omega_0 + \omega_1)} \right) \frac{(\mathbf{n}_0 \cdot \mathbf{n}_1)^{-1} - 1}{(2k-1)\pi} \right), \tag{S28}$$

*with $k \in \mathbb{N}^+$. This equation is valid as long as $B \geq B_{\mathrm{crit}}$, with:*

$$B_{\mathrm{crit}} = \frac{A_{sign(\gamma_n)} + \mathrm{sign}(\gamma_n)\sqrt{A_{\mathrm{sign}(\gamma_n)}^2 - 4 s_0 s_1 (A_\perp^2 + A_\parallel^2)}}{2\gamma_n}, \tag{S29}$$

$$A_{\mathrm{sign}(\gamma_n)} = (s_0 + s_1)A_\parallel + 1.17 \cdot \mathrm{sign}(\gamma_n)|A_\perp|. \tag{S30}$$

### 3. Application to a particular nuclear spin configuration

We now apply the theorem to the nuclear spin configuration identified in this work, i.e., ($A_\parallel = -23.6\,\text{kHz}$, $A_\perp = 12.2\,\text{kHz}$, $\gamma_n = -84.65\,\text{kHz/G}$). For these parameters, we obtain a critical magnetic field strength of $B_{\text{crit}} = 60.5\,\text{G}$ with an $\epsilon_\tau$ of the first resonance of $0.083 < 0.1$.

Comparing the approximated value $\tau_{\text{approx}}^{(1)}$ to the numerically solved equation (S17), we find that the relative error is very small, $(\tau_r - \tau_{\text{approx}}^{(1)})/\tau_r = 0.003$. This stands as a clear confirmation for the quality and validity of our first-order approximation and the criteria used to determine the minimum required magnetic field $B_{\text{crit}}$.

Further considering the magnetic field used for the experiment in this work ($B = 81\,\text{G}$), we find that the first-order approximation describes the CPMG resonance time within an error of 0.0026.

Figure S17 shows the error of the approximation $((\tau_r - \tau_{\text{approx}}^{(1)})/\tau_r)$ for varying magnetic field strengths. The horizontal line represents the maximum acceptable error threshold of 0.003 that is obtained with the constraint $\epsilon_\tau \leq 0.1$. The vertical line represents the critical magnetic field strength ($B_{\text{crit}}$). Importantly, the error is below the threshold for any magnetic field exceeding $B_{\text{crit}}$, thus validating our theorem.

This shows that our analytic expression for determining the resonance times in decoupling sequences can be applied even in low magnetic field conditions in which $|\omega_L| \sim |A_\perp|$.

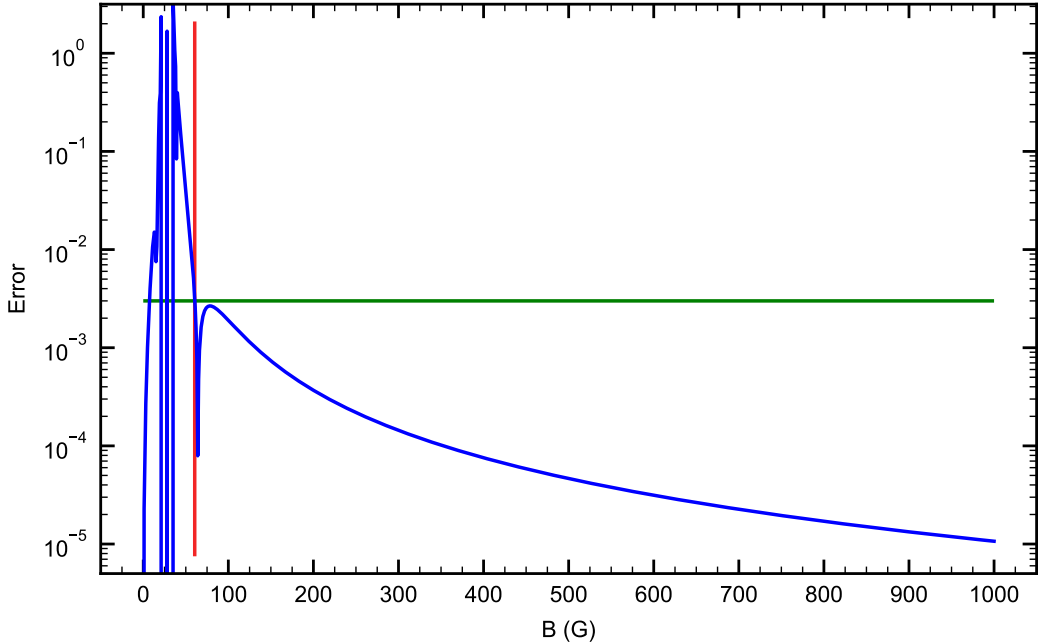

Figure S17. Relative error between the approximated resonance time $\tau_{\text{approx}}$ and the numerical solution of equation (S17). The red line represents the critical magnetic field from which on the error is always below the defined threshold value (represented by the green line).

### D. Obtaining the hyperfine coupling terms

In Figure 2e in the main text, we show the Hahn echo signal at a magnetic field strength of $B = 36\,\mathrm{G}$. We use equation (S12) to fit the signal considering coupling between a single V2 centre with a single nuclear spin. By minimizing the error squares between data and fit, we identify that the nuclear spin must be a $^{29}$Si nucleus, and we obtain initial hyperfine coupling parameters. However, Hahn echo sequences show a limited precision, such that we perform additional measurements. In particular we perform a 8-pulse CPMG sequence at elevated magnetic field ($B = 81\,\mathrm{G}$), see Figure S18. To simulate the signal, we use the probability given by the equation (S7) in the presence of a single nuclear spin. With the hyperfine coupling parameters extracted from the Hahn echo, we simulated the expected 8-pulse CPMG signal. Although the simulated signal shows a good overlap with the data, we find additional dips at $\tau = 3.7\,\mu\mathrm{s}$, $11.1\,\mu\mathrm{s}$ and $18.5\,\mu\mathrm{s}$, which cannot be explained by the presence of a single nuclear spin. We thus attempt to fit the additional dips with a second nuclear spin and obtain a very good qualitative overlap. Subsequently, all four hyperfine coupling terms are fine-tuned to minimize the error squares between the simulation and the experimental data. In the case of several nuclear spin, the probabilities given by the equation (S7) for each nuclear spin are simply multiply as we neglect any nuclear-nuclear interactions. This allows us to precisely determine that there are two $^{29}$Si nuclei with hyperfine coupling terms $A_{\parallel,1} = 2\pi \cdot (-23.5)\,\mathrm{kHz}$, $A_{\perp,1} = 2\pi \cdot 12.0\,\mathrm{kHz}$, $A_{\parallel,2} = 2\pi \cdot 0.2\,\mathrm{kHz}$ and $A_{\perp,2} = 2\pi \cdot 8.5\,\mathrm{kHz}$. To verify our assumption, we simulate, the signal obtained for a varying number of refocusing pulses with $\tau = 5.38\mu\mathrm{s}$. As seen in Figure 2f, the oscillations are perfectly described by our model. As the overlap between data and simulations shows now essentially no residuals, we consider that coupling to additional nuclear spins is very weak compared to the electron spin coherence time, and thus negligible. The system is therefore fully described by one electron and two nuclear spins.

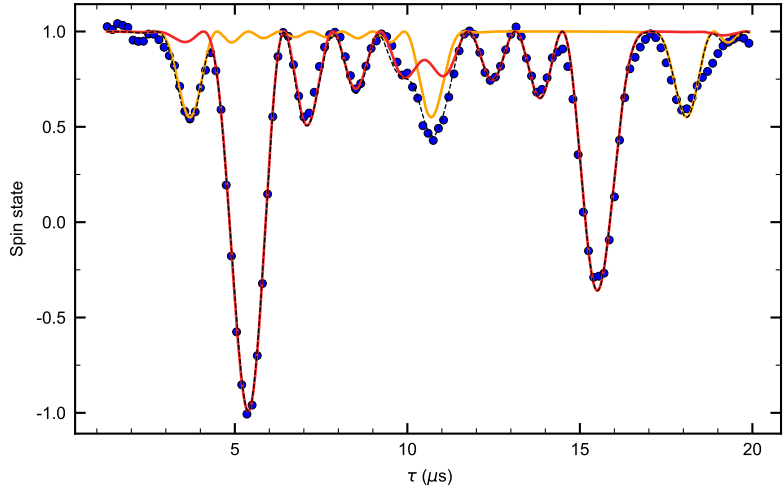

Figure S18. Spin signal for the 8-pulse CPMG sequence at $B = 81\,\mathrm{G}$ (blue dots) and the corresponding contribution of each nuclear spins (red and orange lines). The fit to the data (black dashed line) is based on the product of both contributions.

## E. Gate fidelity

In this section, we simulate the evolution of the nuclear spin for decoupling sequences with $\tau = \tau_r$ and varying number of pulses $N$. From here, we determine the nuclear spin rotation fidelities with respect to relevant quantum gates, such as Hadamard, NOT and I. We use equation S7 and express the evolution operator $\hat{\mathcal{U}}_i(\tau)$ as:

$$\hat{\mathcal{U}}_i(\tau) = \cos\left(\frac{\omega_i \tau}{2}\right)\hat{\mathbb{1}} - i \cdot \sin\left(\frac{\omega_i \tau}{2}\right)\mathbf{n}_i \cdot \hat{\mathbf{I}}. \tag{S31}$$

The nuclear spin evolution is simulated with the hyperfine parameters obtained from the fit to the 8-pulse CPMG measurements ($A_\parallel = -23.5\,\text{kHz}$ and $A_\perp = 12.0\,\text{kHz}$ for the first nuclear spin). For $N = 4$, we find that the rotation axes and rotation angles for the operators $\hat{U}$ and $\hat{V}$ are:

$$\mathbf{n}_{\hat{U}} \approx \sqrt{0.943} \cdot \mathbf{e_x} + \sqrt{0.057} \cdot \mathbf{e_z}, \tag{S32}$$

$$\mathbf{n}_{\hat{V}} \approx \sqrt{0.941} \cdot \mathbf{e_x} - \sqrt{0.059} \cdot \mathbf{e_z}, \tag{S33}$$

$$\theta_{\hat{U}} = \theta_{\hat{V}} \approx 0.49 \cdot \pi. \tag{S34}$$

We make the following observations:

- The rotation axes are essentially perfectly anti-parallel: $\mathbf{n}_{\hat{U}} \cdot \mathbf{n}_{\hat{V}} = -0.999994$.

- The rotation angles are the same.

- The axis of rotation are (almost entirely) oriented along the $\mathbf{e_x}$ axis.

From the simulation, we now infer the gate fidelities for $N = 4, 8, 16$. To do so, we initialize the bipartite system in one of the four spin-eigenstates, and simulate the spin state $|\Psi_f\rangle$ at the end of the sequence. We then calculate the overlap with the target spin-state as $\mathcal{F} = |\langle\Psi_{\text{target}}|\Psi_f\rangle|^2$. For particular quantum gates, we obtain the following fidelities:

- For $N = 4$, the fidelity for the creation of the Bell state $|\Phi_i^\pm\rangle$ and $|\Psi_i^\pm\rangle$ is $\mathcal{F} = 97\%$.

- For $N = 8$, the fidelity for performing the X-gate on the nuclear spin is $\mathcal{F} = 94\%$.

- For $N = 16$, the fidelity for performing the identity-gate on the bi-partite system is $\mathcal{F} = 98\%$.

We mention that the fidelities could be further improved using an increased external magnetic field. This would further suppress spurious rotation around the $\mathbf{e_x}$ axis. In addition, the nuclear spin rotation angle per CPMG refocussing pulse would decrease, which would allow to precisely tune the number of pulses to implement particular gates. Obviously, the price to pay would be an increased number of refocussing pulses, as well as increased total gate duration times.

---

* c.babin@pi3.uni-stuttgart.de

† f.kaiser@pi3.uni-stuttgart.de